\documentclass[a4paper,fleqn,usenatbib]{mnras}

\usepackage{newtxtext,newtxmath}

\usepackage[T1]{fontenc}
\usepackage{ae,aecompl}

\usepackage{graphicx}	

\newcommand{\ceio}{C$^{18}$O }
\newcommand{\twco}{$^{12}$CO }
\newcommand{\thco}{$^{13}$CO }
\newcommand{\kms}{km\,s$^{-1}$}

\title[Dynamics of the Western Wall]{Length-scales and Dynamics of Carina's Western Wall}

\author[T. P. Downes et al]{
Turlough P. Downes$^1$\thanks{E-mail: turlough.downes@dcu.ie (TPD)}, Patrick Hartigan$^2$, Andrea
	Isella$^2$
\\
$^1$Centre for Astrophysics \& Relativity, School of Mathematical Sciences, Glasnevin, Dublin 9,
		 Ireland. \\
$^2$Department of Physics and Astronomy, Rice University, 6100 Main Street, MS 108, Houston, TX
77005, USA.
}

\date{Accepted XXX. Received YYY; in original form ZZZ}

\pubyear{2020}

\begin{document}
\label{firstpage}
\pagerange{\pageref{firstpage}--\pageref{lastpage}}
\maketitle

\begin{abstract}
We present a variety of analyses of the turbulent dynamics of the boundary of a photo-dissociation region (PDR) in the 
Carina Nebula using high resolution ALMA observations. Using Principal Component Analysis we
suggest that the turbulence in this molecular cloud is driven at large scales. Analysis of the centroid velocity structure
functions indicate that the turbulence is dominated by shocks rather than local (in $k$-space) transport of energy. We further find 
that length-scales in the range 0.02 -- 0.03\,pc are important in the dynamics of this cloud and this finding is supported by analysis 
of the dominant emission structure length-scale. These length-scales are well resolved by the observational data and we conclude that the 
apparent importance of this range of scales is physical in origin. Given that it is also well within the range strongly influenced by 
ambipolar diffusion, we conclude that it is not primarily a product of turbulence alone, but is more likely to be a result of the 
interplay between gravity and turbulence. Finally, through comparison of these results with previous observations of H$_2$ emission
from the Western Wall we demonstrate that observations of a PDR can be used to probe the internal structure of the undisturbed portion of
a molecular cloud.
\end{abstract}

\begin{keywords}
ISM:kinematics and dynamics -- ISM: magnetic fields -- turbulence -- methods: statistical
\end{keywords}



\section{Introduction}

Turbulence in molecular clouds is known to be likely to have an impact on star formation \cite[e.g.][]{krumholz05, kainulainen13, burkhart15, kainulainenfed17}, and 
even planet formation. Much of this turbulence is thought to be driven by large-scale flows, with the potential for smaller scale turbulence also excited
by stellar feedback. The energy budget of this turbulence must be large as it is both supersonic and fast-decaying \citep{stone_etal98, maclow_etal98} and 
understanding it should give us insight into the origin of the Initial Mass Function \citep[e.g.][]{nam_etal21}, molecular cloud formation and galactic dynamics. 
The nature of the turbulent cascade is of particular interest, both from a theoretical perspective and also a more pragmatic one, since the properties of this 
cascade are what create the initial conditions from which stars and planets ultimately form.

There is a large body of literature on the theory \cite[e.g.][among many others]{choetal09, lazetal12, ballesteros-paredesetal18, 
xulaz20b, xulaz20a}, numerical simulations \citep[][for example]{ossenkopf02, downes12, burkhart15, beattieetal19, wollenbergetal20} and 
observations \citep[for example][]{crutcher99, abel06, hennebelle12, sun20} of interstellar turbulence. One of the primary challenges in 
relating observations to numerical simulations and theory is that there is considerable difficulty in extracting turbulence characteristics 
from observations. The light we observe from these regions is a convolution of the effects of density (emission and 
absorption), temperature, line-of-sight projection and velocity. Projection can be expected to average out velocity and density 
variations along the line of sight through so-called ``projection smoothing''. This happens preferentially for variations which occur on 
length-scales short in comparison to the thickness of the cloud being observed. As a result this projection smoothing can yield misleading 
interpretations of how the turbulent energy cascade proceeds. In addition if large-scale magnetic fields are dynamically important the turbulence can 
be anisotropic \citep[e.g.][]{choetal_02, beattieetal_20}, as would be expected from the theory of \citet{gs95}: then the angle between the line-of-sight and the 
large-scale field becomes a critical component in determining what we actually observe. Indeed, while on the largest turbulence scales this may not 
be an issue, it will always be an issue below the length-scale at which the turbulent velocity is of order the Alfv\'en speed.

Notwithstanding these difficulties, in recent years significant progress has been made in understanding how we can relate 
observations of turbulent regions to the nature of the turbulence in those regions \citep[see, e.g.][]{stewart_22}. Principal Component Analysis (PCA) can be 
used as a tool for determining the length-scales on which turbulence is driven \citep{brunt_etal_09}, while both PCA and the centroid 
velocity structure functions have also been used to great effect in determining the nature of the turbulence cascade 
\citep{brunt_etal03}. In particular, these structure functions can be used to distinguish between Kolmogorov-type turbulence where 
energy is transferred by local interactions in $k$-space and shock-dominated turbulence in which the energy is transferred 
directly from large scale motions to the diffusion scale. It is worth keeping in mind that, even in supersonic turbulence, there is a length-scale below which relative motions
will be subsonic/sub-Alfv\'enic. The scale is called the ``sonic scale'' and we expect a Kolmogorov-like cascade below the sonic scale 
in all cases. In any case, the different energy cascades at larger scales produce 
differing density and magnetic field distributions, thereby setting different initial conditions for star and planet formation \citep[e.g.][]{vazquez_03, federrath21}. It is of interest to note
that the diffusion scale in molecular cloud turbulence is not a single, well-defined scale. Diffusion processes occur both
as a result of viscosity, and of non-ideal magnetohydrodynamic effects. These latter effects are believed to be important
on length-scales up to at least 0.05\,pc \citep{downes12}.

Once stars begin to form, of course, they may begin to influence their surroundings and thus the formation
of other stars in their neighbourhood \citep[e.g.][]{menonetal_21}. One of the relatively close, but extreme, sites of star formation is the Carina 
Nebula. Radiation from the Carina OB1 association impinges on a nearby molecular cloud, forming a photo-dissociation
region (PDR) at its boundary. Recently, ALMA observations were carried out on a small section of this PDR
which incorporated part of the body of the molecular cloud itself. The unprecedented angular resolution of ALMA allows
the resolution of structures as small as 0.01\,pc \citep{hartigan2022} which is significantly smaller than the length-scale at
which ambipolar diffusion can be expected to impact the turbulent cascade. 

In this paper we investigate the turbulence characteristics of this molecular cloud, focusing particularly on finding
any notable length-scales, with a view to understanding the dynamics occurring within the cloud in a statistical sense.
In Section \ref{sec:obs} we briefly outline the observations upon which we base our analysis. Section \ref{sec:analysis}
contains a detailed description of the analyses carried out, while Sect.\ \ref{sec:results} contains the results of
these analyses. Finally, we discuss our conclusions in Sect.\ \ref{sec:conclusion}.

\section{Observations}
\label{sec:obs}


We give a brief summary of the acquisition and processing of the ALMA data analysed in this work and refer the reader to \citet{hartigan2022} for 
further details.  We used the Atacama Large Millimeter/Submillimeter Array (ALMA) to map a section of Carina's Western Wall. This is a bright 
PDR arising from radiation emitted by O-stars from the open clusters Trumpler 14 and Trumpler 16 which are nearby. The maps are 
of a region approximately 0.7\,pc\,$\times$\,0.9\,pc in size, assuming a distance of 2.3\,kpc. The J=2-1 transitions of \twco, \thco and \ceio were 
recorded by tuning ALMA's receivers to 1.3\,mm. Observations of [CI] 609\,$\mu$m were also performed and are detailed in \cite{hartigan2022}. Each line
was recorded using channels with a velocity resolution of 0.166 \kms. After processing the final spatial resolution of the maps was approximately
1'', or 0.011\,pc at 2.3\,kpc. The line emission from \twco and \thco was found to have optical depths greater than 1, while \ceio was found to
have optical depths of around 0.5.

\section{Analysis}
\label{sec:analysis}
In this section we outline the analyses undertaken. Firstly, we describe a Principal Component Analysis, PCA,
following \citet{brunt_etal_09} and \citet{brunt_heyer_13}. The primary aim of this analysis is
to determine the driving scale of turbulence in the Western Wall. Secondly, we perform an
analysis of the spatial distribution of the emission to determine whether there are preferred
length-scales in the observations. Finally, we perform an analysis of the structure functions of
the line-of-sight velocity field to compare with simulation results in the literature, such as
those of \citet{downes12} and \citet{federrath16a}.

All analyses were carried out for each isotope observed. We performed a masking process whereby pixels with a value less 
than a certain threshold were ignored. Unless otherwise specified this masking was performed at the level of a factor of 
five times the RMS noise in the data. For the \twco the rms noise was measured to be 6.5\,mJy/Beam, for \thco it was 
6\,mJy/Beam and for \ceio it was 4.5\,mJy/Beam.

\subsection{Principal Component Analysis}
\label{sec:pca_analysis}
Given a multivariate dataset, the underlying goal of PCA is to
identify which variable or, more generally, which combinations of variables are responsible for
variations in the data. In our case the data in question is the recorded intensity, while the variables are the position and the line-of-sight velocity.
Briefly, we can represent our data cube by a matrix, $\mathbf{A} = (a_{ij})$, where $a_{ij} = I(v_i, \mathbf{r}_j)$ 
where $I(v_i, \mathbf{r}_j)$ is the intensity in the i$^{\rm th}$ velocity (spectral) bin and $\mathbf{r}_j$ is the position in the image. Note that $\mathbf{r}_j$ should
be viewed as a 1D array of position vectors defining all locations (pixels) in the image. We then
calculate our covariance matrix by $\mathbf{C} = (c_{jk}) = (a_{ij} a_{ik})$, adopting the usual Einstein notation. The associated set of eigenvalue 
and eigenvector pairs yields information on the variation in the data. Typically these pairs are ordered by the size of the eigenvalues, and then the
eigenvectors give the combinations of independent variables responsible for the variation in the data, ordered from greatest variation to least variation. We 
can create the associated set of eigenimages by taking the inner product of each eigenvector with $\mathbf{A}$. The ratios of the length-scales in these ordered
eigenimages, usually defined to be that length at which the autocorrelation function of the eigenimage has dropped to $1/e$ of its maximum, has been shown to 
correlate well with the driving scales of any turbulence present in the cloud \citep{brunt_heyer_13}.

In this work we apply this analysis to the \twco, \thco and \ceio data cubes with a view to determining whether
the turbulence present is likely to be driven at scales larger than the Western Wall itself. Our conclusions are restricted in the 
sense that, if PCA indicates that turbulence is being driven at large scales then it is in principal possible that
the turbulence is being driven at scales larger than the analysed subsets of the data (see Sect.\ \ref{sec:analysed-regions}), but smaller than our total observed 
field. However, our subsets are sufficiently large that this is unlikely to impact our conclusions qualitatively.

\subsection{Centroid velocity structure functions}

Another, complementary, approach to studying the properties of turbulence in molecular cloud observations is to investigate the centroid
velocity structure functions \citep{hily-blantetal_08, federrathetal_10, downes12}.  We use these functions to study the length-scales associated with variations in the 
velocities, as distinct from length-scales associated with variations in the intensity of the emission although the two cannot be completely deconvolved, of course. 
The properties of these functions also yield clear information regarding the nature of the energy cascade, in particular the dependence of the turbulent energy 
density on length-scale.

We construct the centroid velocity structure functions of orders 1 - 3 from our observations to compare with simulation 
results and theory. First we define our centroid velocity to be
\begin{equation}
v_{\rm c}(\mathbf{r}_j) = \frac{\sum_{i} v_i I(v_i, \mathbf{r}_j)}{\sum_i I(v_i, \mathbf{r}_j)}
\end{equation}
We can then introduce the quantity
\begin{equation}
S_{\ell}(\mathbf{r}_j,r) = \left<\left|v_{\rm c}(\mathbf{r}_j) - v_{\rm c}(\mathbf{r}_k)
	\right|^{\ell}\right>_ {|\mathbf{r}_j - \mathbf{r}_k| = r}
\end{equation}
where the angle brackets indicate averaging over all $\mathbf{r}_k$ for which $|\mathbf{r}_j
- \mathbf{r}_k| = r$.  Given that we have two (spatial) dimensional data this corresponds to averaging over a circle
of radius $r$ centred on $\mathbf{r}_j$. If $\ell=1$ then $S(\mathbf{r}_j, r)$ is the average of the difference in the centroid velocity at 
$\mathbf{r}_j$ and the velocity at all other points a distance $r$ from $\mathbf{r}_j$. The centroid velocity structure function of 
order $\ell$ is then defined by
\begin{equation}
C_{\ell}(r) = \left<S(\mathbf{r}_j, r)\right>_{\mathbf{r}_j}
\end{equation}
\noindent where the angle brackets indicate averaging over all $\mathbf{r}_j$. For example,
$C_1(r)$ is then the mean difference in velocity between any two points a distance $r$
apart. The results of this analysis are presented in Sect.\ \ref{sec:structure-fn-results}.

\subsection{Emission length-scale analysis}
\label{sec:emission-lengthscale-analysis}

To discover whether there are dominant length-scales in the emission structures, rather than the 
velocities, we proceed as follows. For each of the velocity channels associated with the Western Wall we performed
standard edge detection analyses using a variety of techniques including the magnitude of the Laplacian of the emission, and the ``edge dog'' 
algorithm using IDL version 8.7.1. We then calculate the wavelet power spectrum of the rows and columns of the resulting image using Morlet wavelets of order 3, and taking into 
account the bias rectification necessary to compare power at different scales/locations. This latter consideration simply entails dividing the power at each 
length-scale by that length-scale. The average of the power spectra for all rows and columns is calculated, yielding a different power spectrum for each velocity 
channel. These power spectra are averaged to produce an overall power spectrum for the data cube. The results of this analysis are presented in 
Sect.\ \ref{sec:emission-length-scales}.

Several wavelets were investigated, together with several edge-finding algorithms and the results presented, in terms of the length-scales at which the peak power
is located, remains the same.

\subsection{Regions analysed}
\label{sec:analysed-regions}

The observed field includes large regions both within and exterior to the Western Wall. Therefore it would not be helpful 
to apply the analyses described to the whole field of view. We choose the regions analysed as follows. Boxes 1, 2 and 3 (see Fig \ref{fig:pca-region})
are each of size $0.11\times0.11$\,pc$^2$ (assuming a distance of 2.3\,kpc), while Box 4 is approximately $0.23 \times 0.23$\,pc$^{2}$. All regions 
were defined to be squares in order to avoid directional bias in the analyses. We now describe the motivation for choosing these regions to analyse:
\begin{itemize}
\item[{\bf Box 1}] South East corner is located at $\alpha$ (2000) = 10:43:31.39 $\delta$ (2000) = -59:36:07.05 and this box is within a region 
reminiscent of the profile of a ``witch's head''. This region is chosen as it may be qualitatively different 
to the rest of the Western Wall: it may be that this region is a protrusion from the Western Wall which may yield different observational statistics.
\item[{\bf Box 2}] South East corner is located at $\alpha$ (2000) = 10:43:29.72 $\delta$ (2000) = -59:36:05.65. Box 2 is located just to the West of Box 1 and 
is chosen as it is within the Western Wall, but not overlapping with any part of the ``witch's head''.
\item[{\bf Box 3}] South East corner is located at $\alpha$ (2000) = 10:43:33.96 $\delta$ (2000) = -59:36:43.35. Box 3 is the same size as Boxes 1 and 2, but 
located in Cloud B which is not so strongly irradiated. This was analysed in order to find if there were any clear differences between the observations of the strongly 
irradiated Western Wall and and less strongly irradiated clouds.
\item[{\bf Box 4}] With its South East corner located at $\alpha$ (2000) = 10:43:29.95 $\delta$ (2000) = -59:36:11.85 this is a much larger region than the other 
Boxes and is chosen to be the largest square area in the observed part of the Western Wall which does not incorporate any part of the witch's head. This region is 
analysed in order to allow us to gain insight into processes operating at larger length-scales.
\end{itemize}
For each box each analysis was carried out on these regions, and on a total of 31 further regions of the same size with their South Eastern corner shifted
by 2 pixels. Where average data is presented the averaging is done over these regions.

%
%

\section{Results}
\label{sec:results}

We first describe the results of the PCA analysis of the observed data for each observed CO isotopologue, then the centroid velocity structure functions
followed finally by an analysis of the dominant length-scales present in the emission.

\subsection{PCA Results}
\label{pca-results}

\begin{figure*}
\centering
\begin{tabular}{cc}
\includegraphics[width=6.5cm]{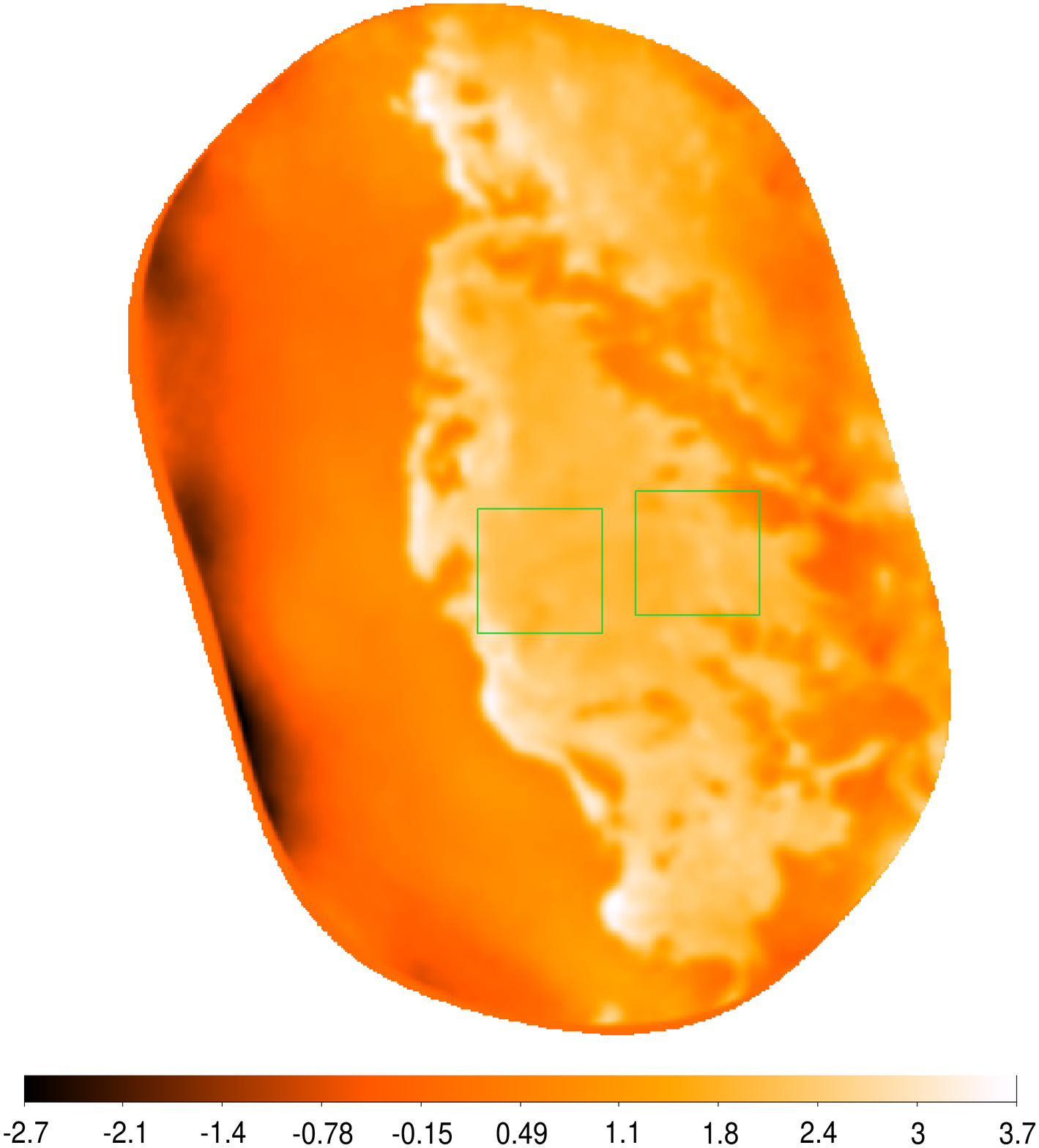} & \includegraphics[width=6.5cm]{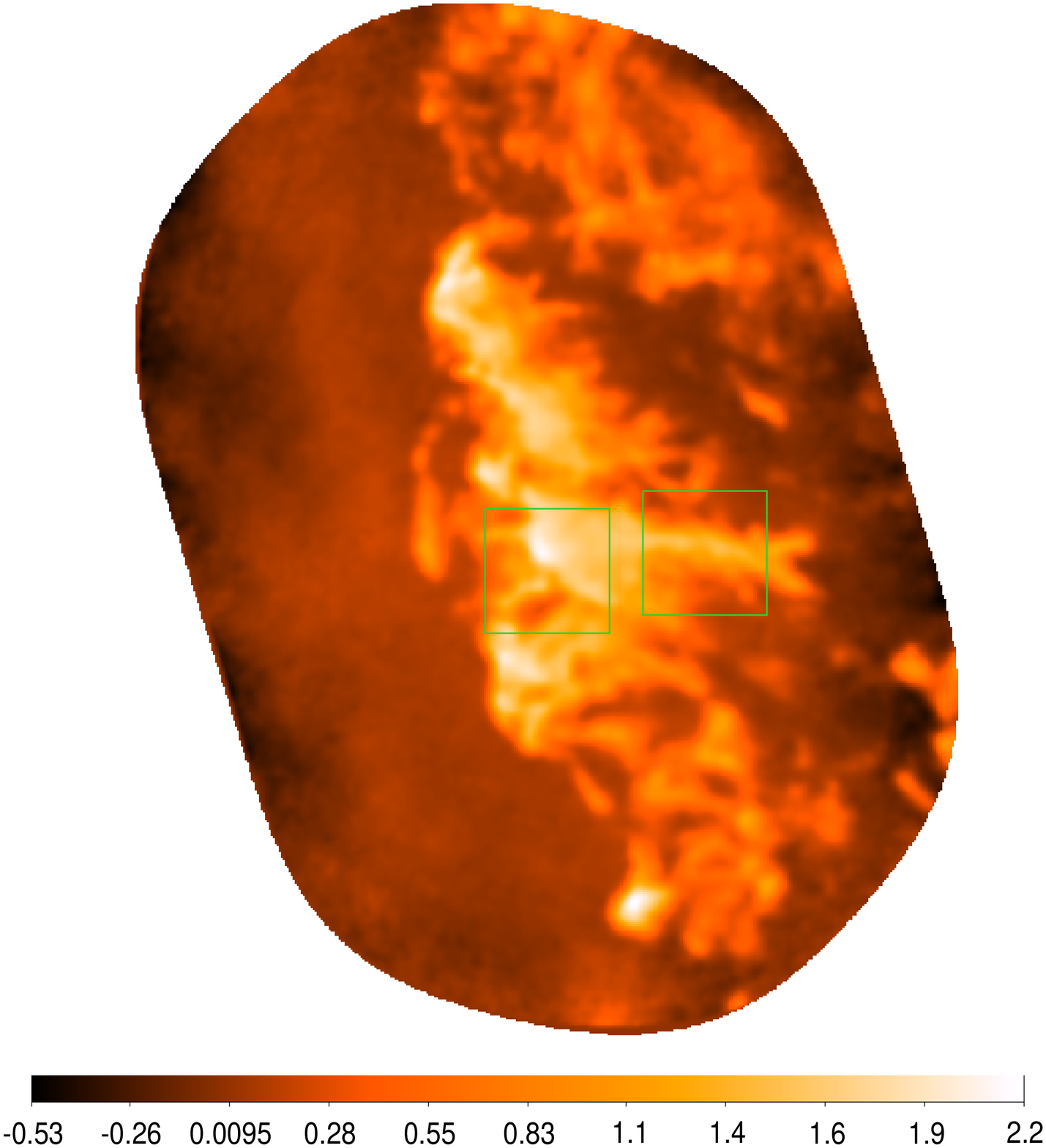} \\
\includegraphics[width=6.5cm]{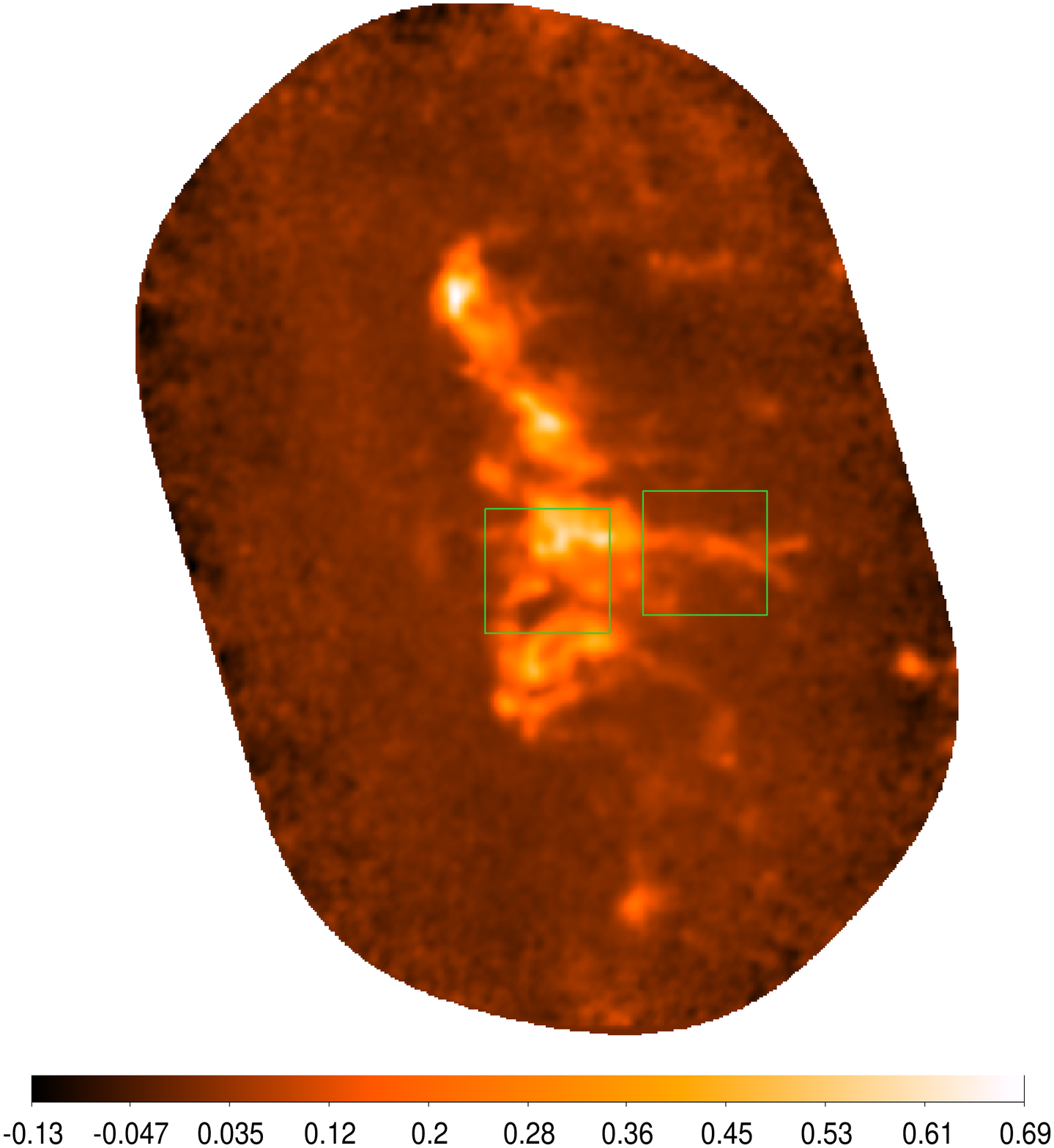} & \includegraphics[width=6.5cm]{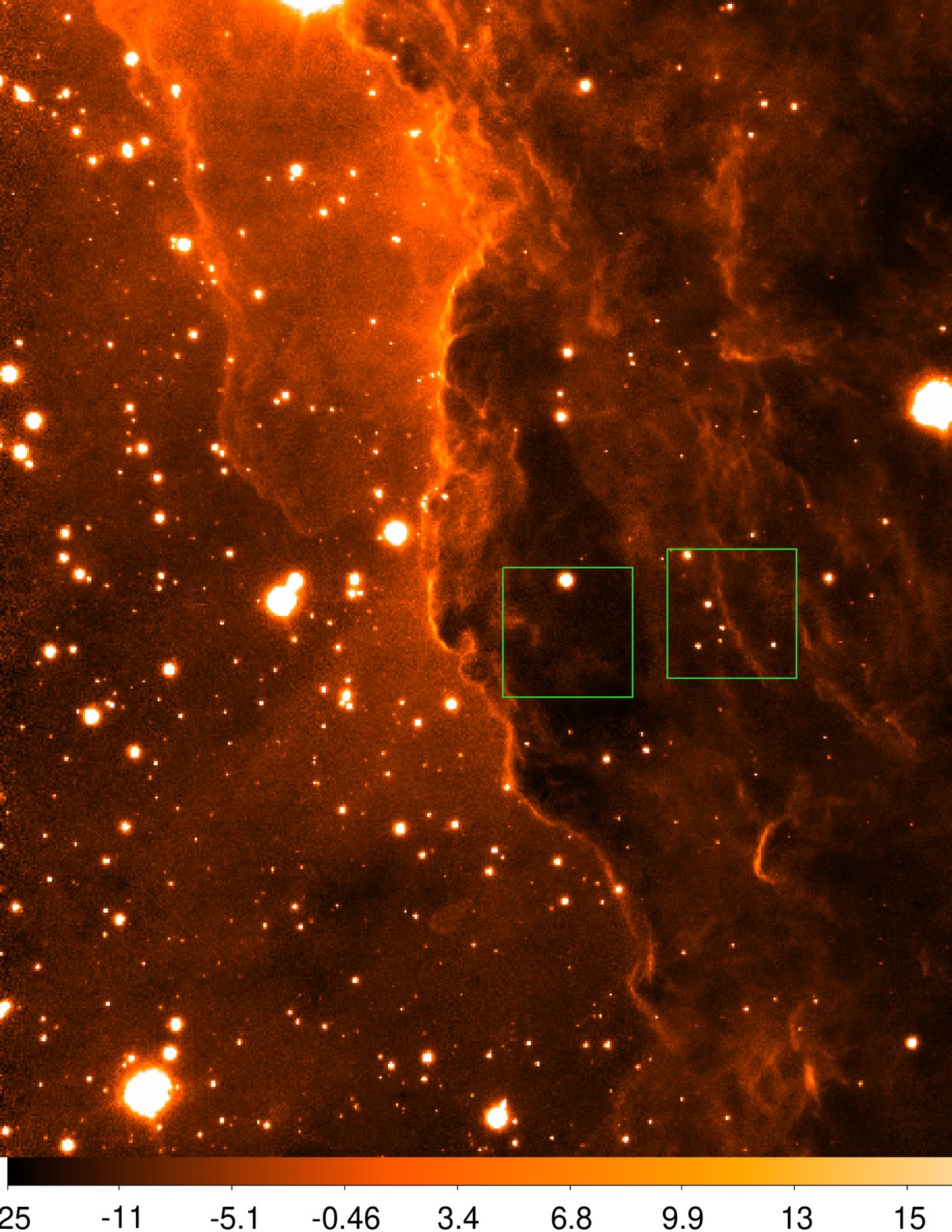} \\
\includegraphics[width=6.5cm]{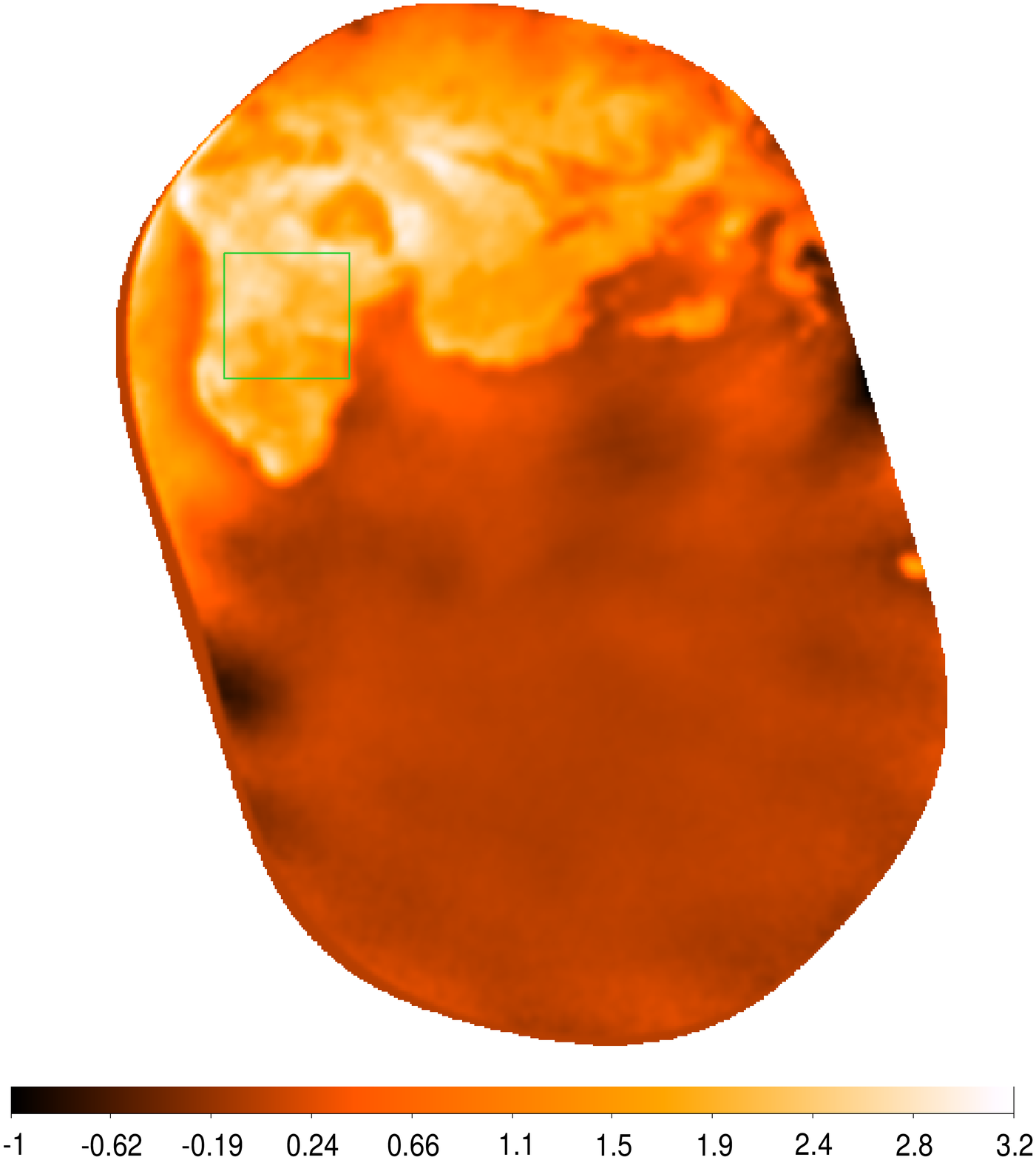} & \includegraphics[width=6.5cm]{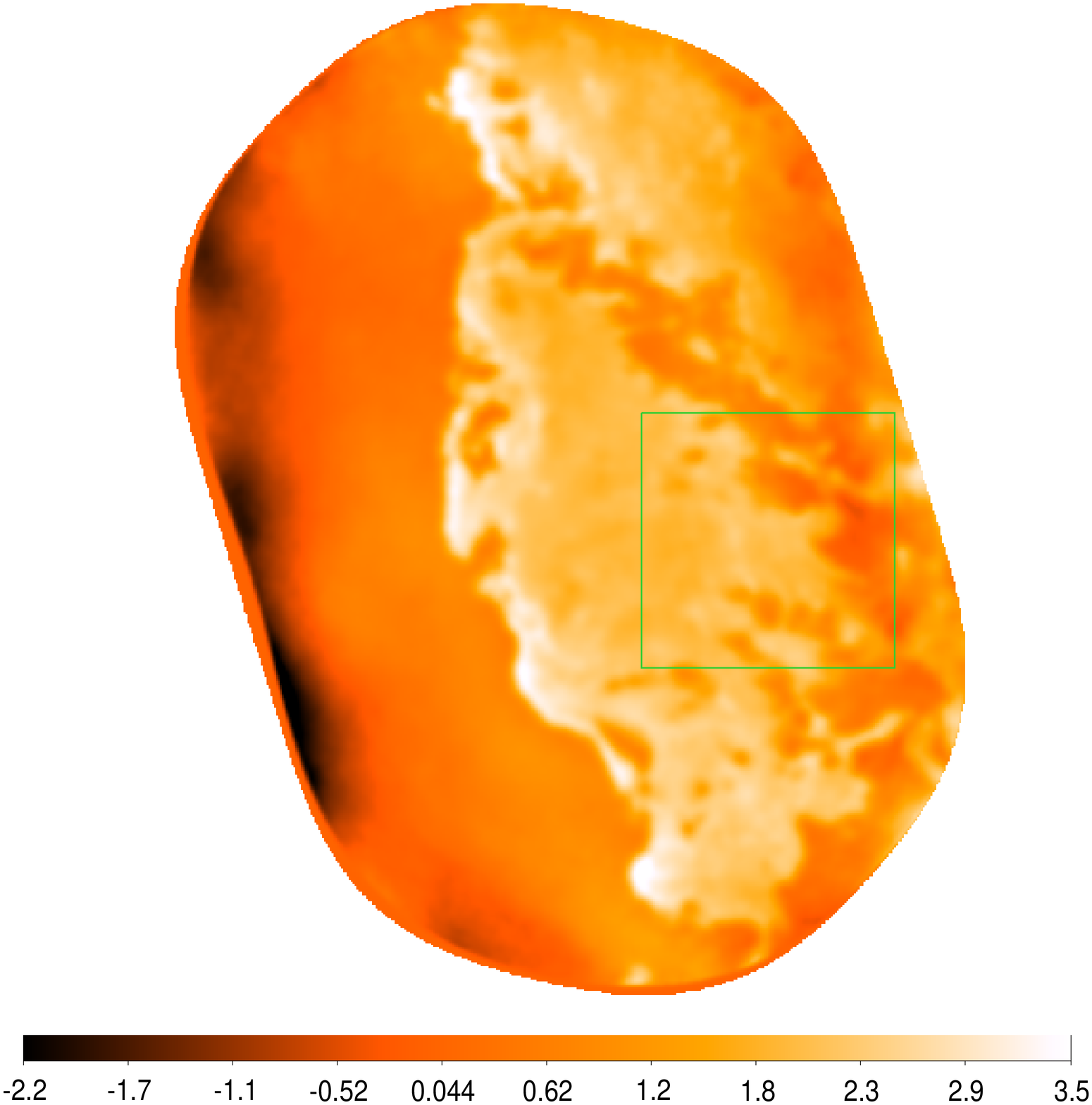}
\end{tabular}
\caption{\label{fig:pca-region} The \twco (top left), \thco (top right), \ceio (middle left) emission at a velocity of -25.27 \kms in the Western 
Wall. The middle right panel contains the H$_2$ image from \citet{hartigan2020} for comparison. The green boxes indicate the regions in which the 
analyses described in the text have been carried out: Box 1 is the left-hand box, while Box 2 is the right-hand box. Finally, the bottom left panel is
the \twco emission for Cloud B \citep{hartigan2022} at a velocity of -12.17 \kms showing the location of Box 3, while the bottom right panel is the 
\twco emission at a velocity of -25.27 \kms showing the location of Box 4.
}
\end{figure*}


The regions defined in Sect.\ \ref{sec:analysed-regions} were all analysed as described in Sect.\ \ref{sec:pca_analysis} using PCA.

\begin{figure*}
\centering
\begin{tabular}{ccc}
\includegraphics[width=4.8cm]{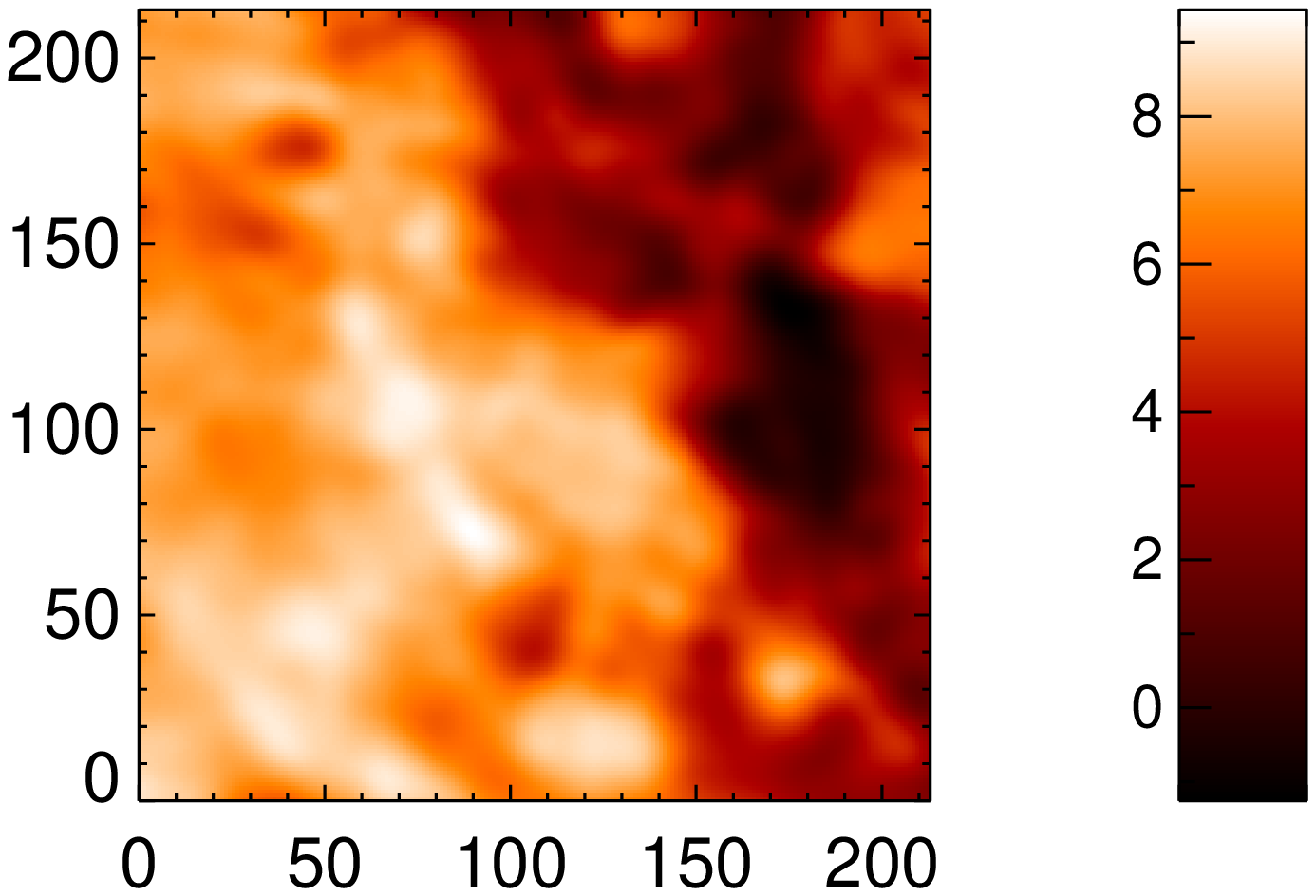} &
\includegraphics[width=4.8cm]{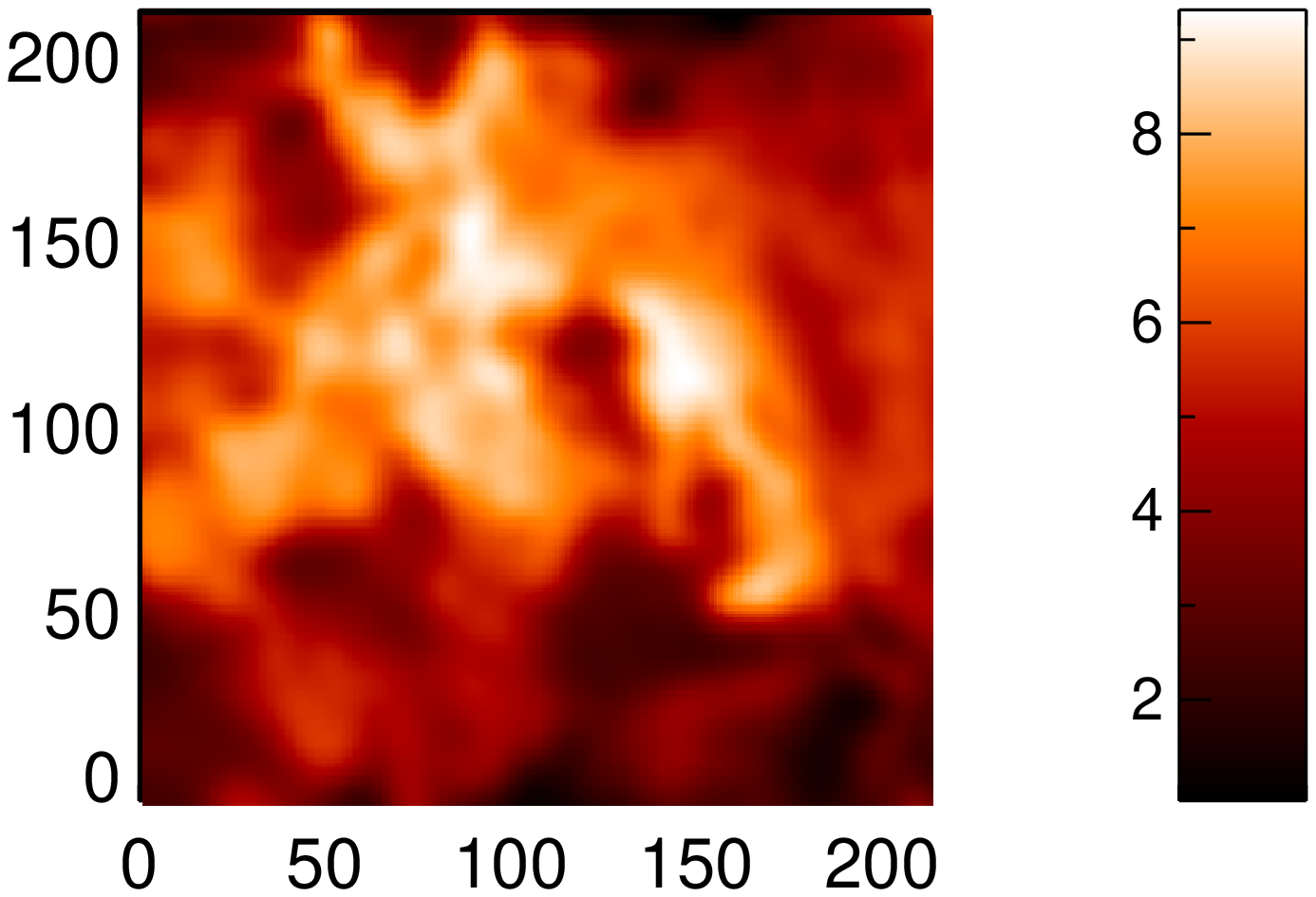} &
\includegraphics[width=4.8cm]{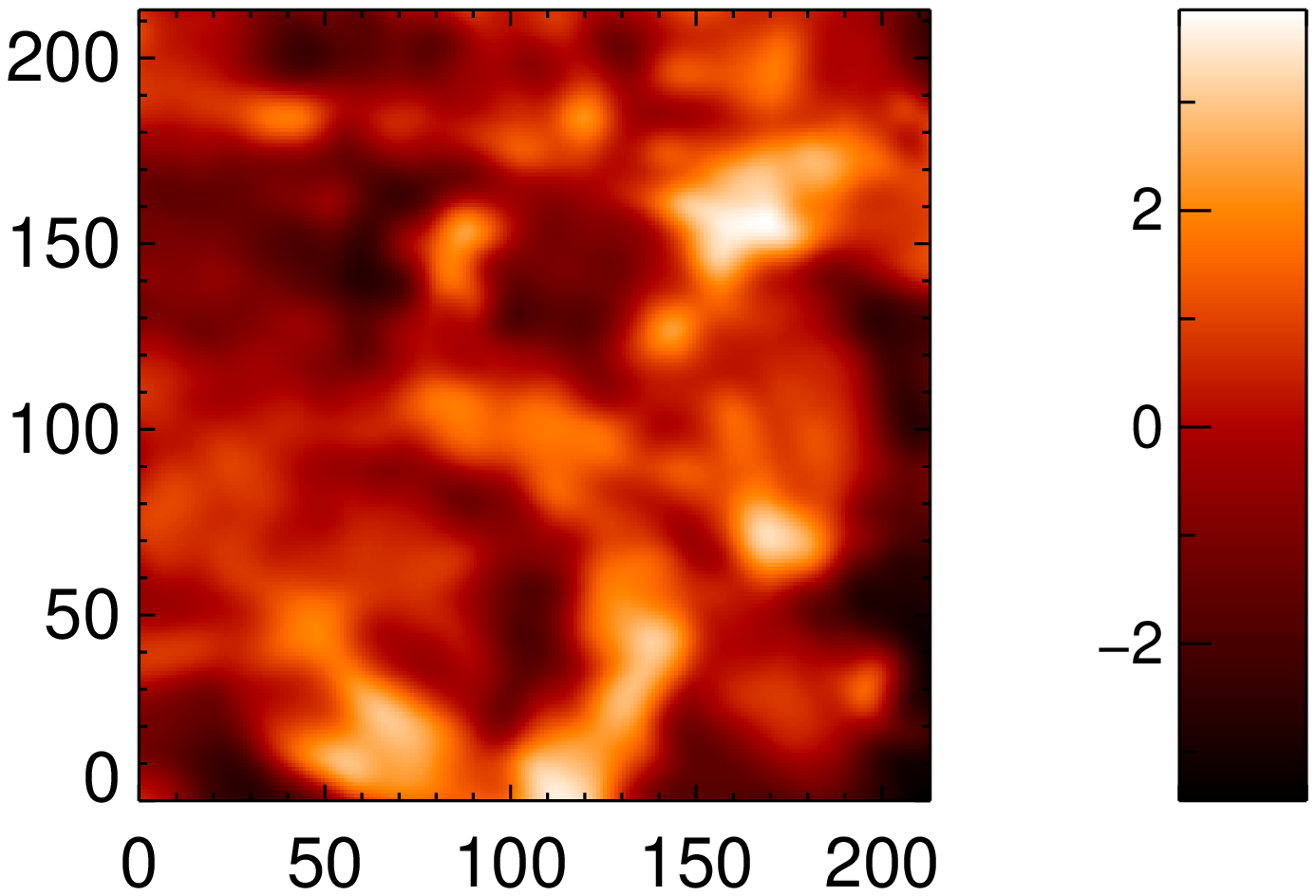} \\
\includegraphics[width=4.8cm]{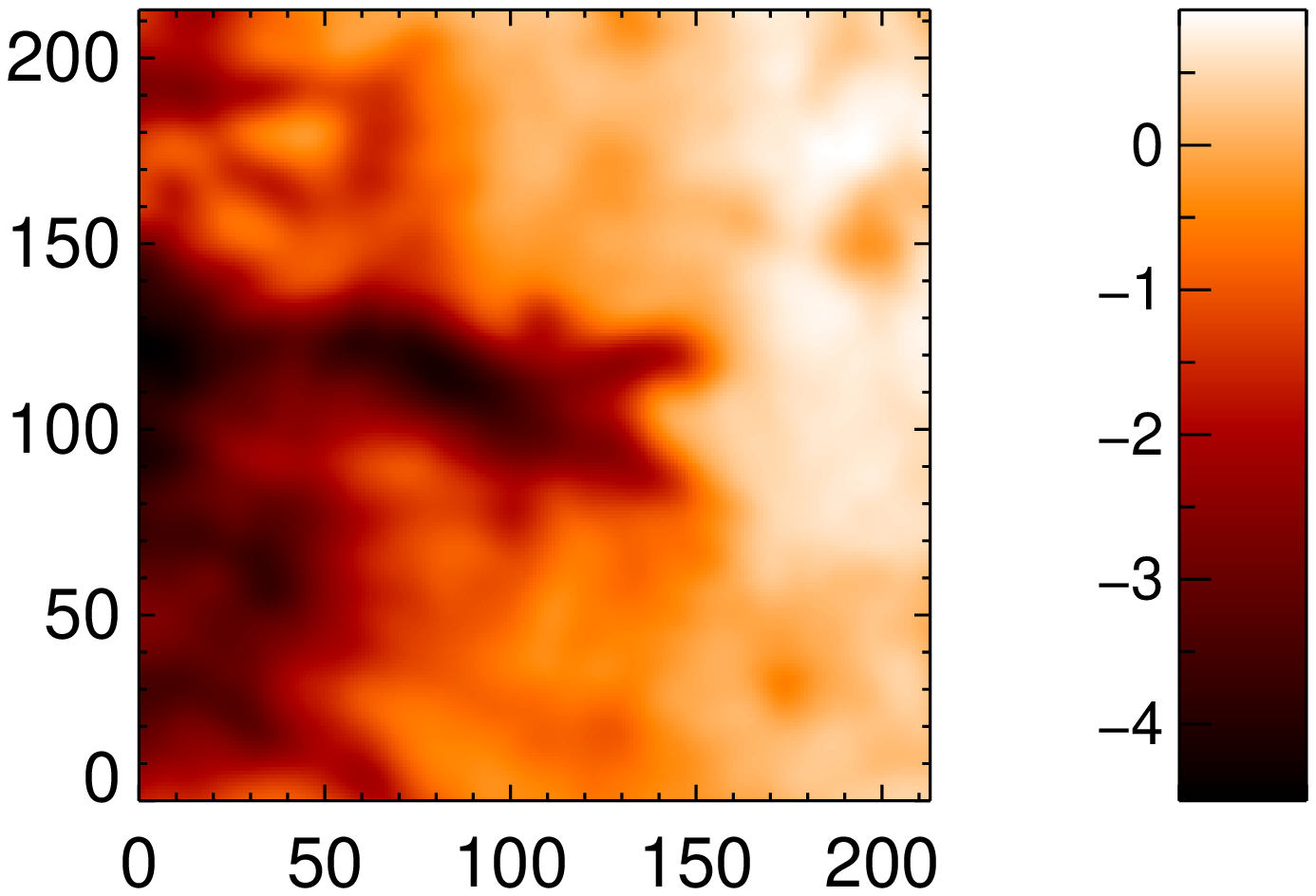} &
\includegraphics[width=4.8cm]{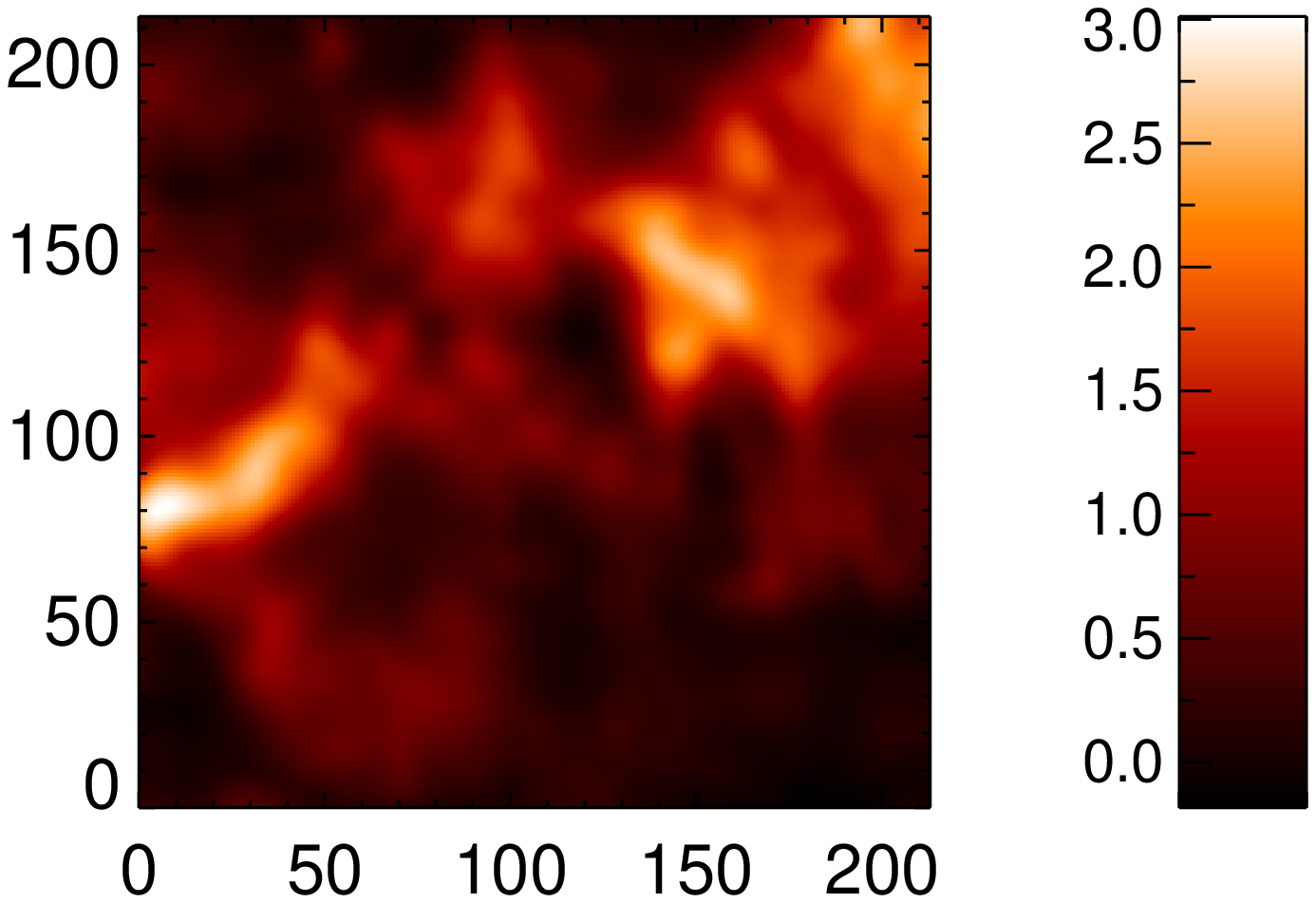} &
\includegraphics[width=4.8cm]{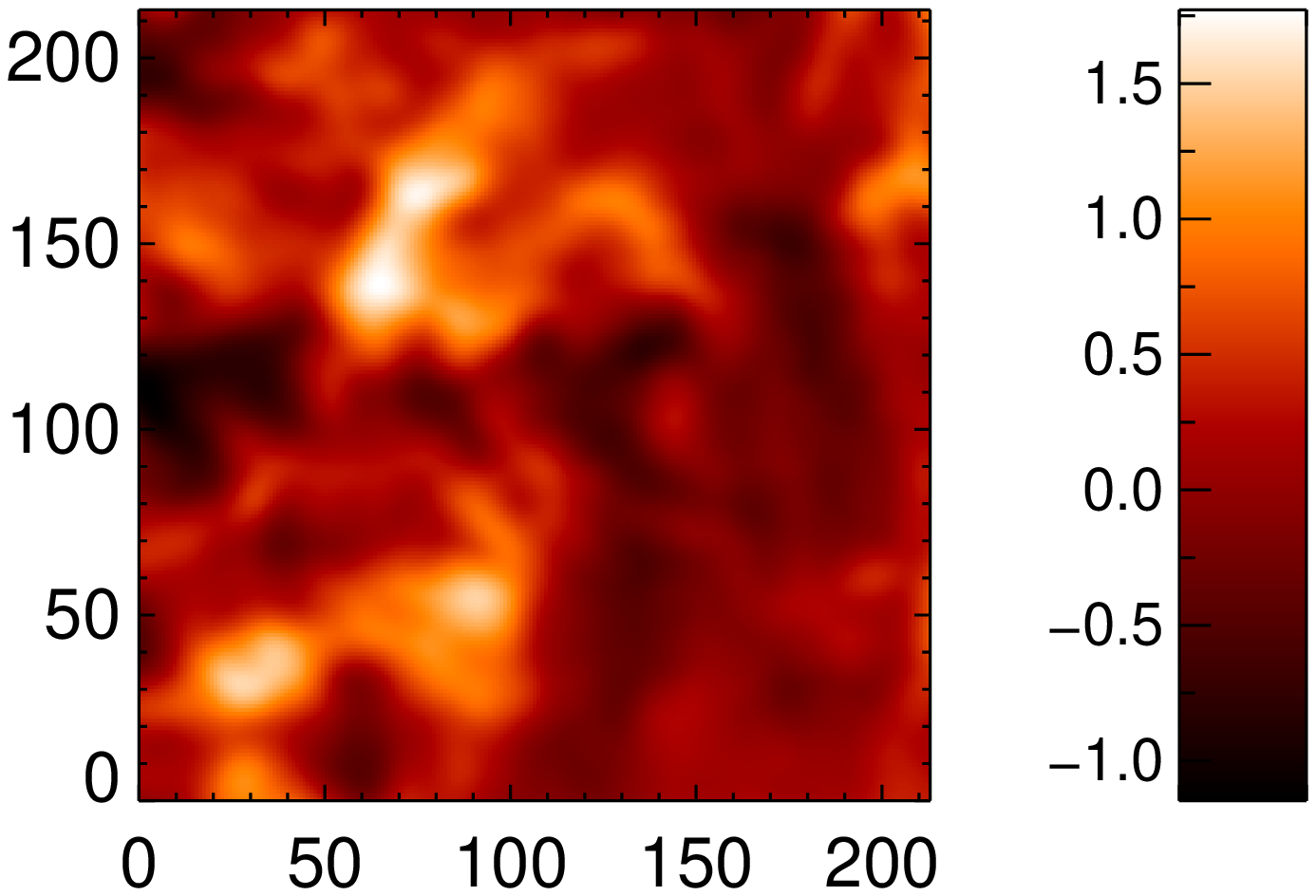} \\
\includegraphics[width=4.8cm]{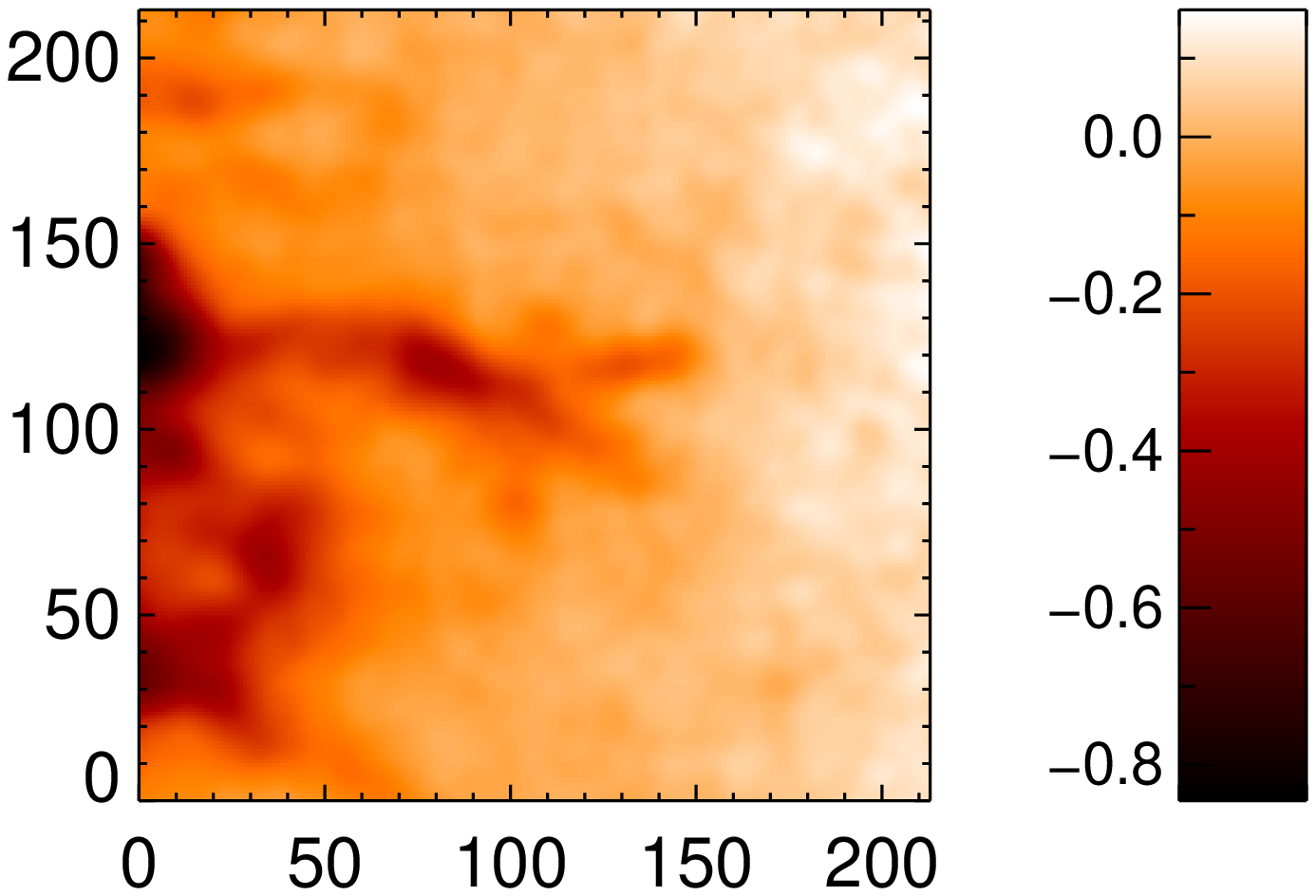} &
\includegraphics[width=4.8cm]{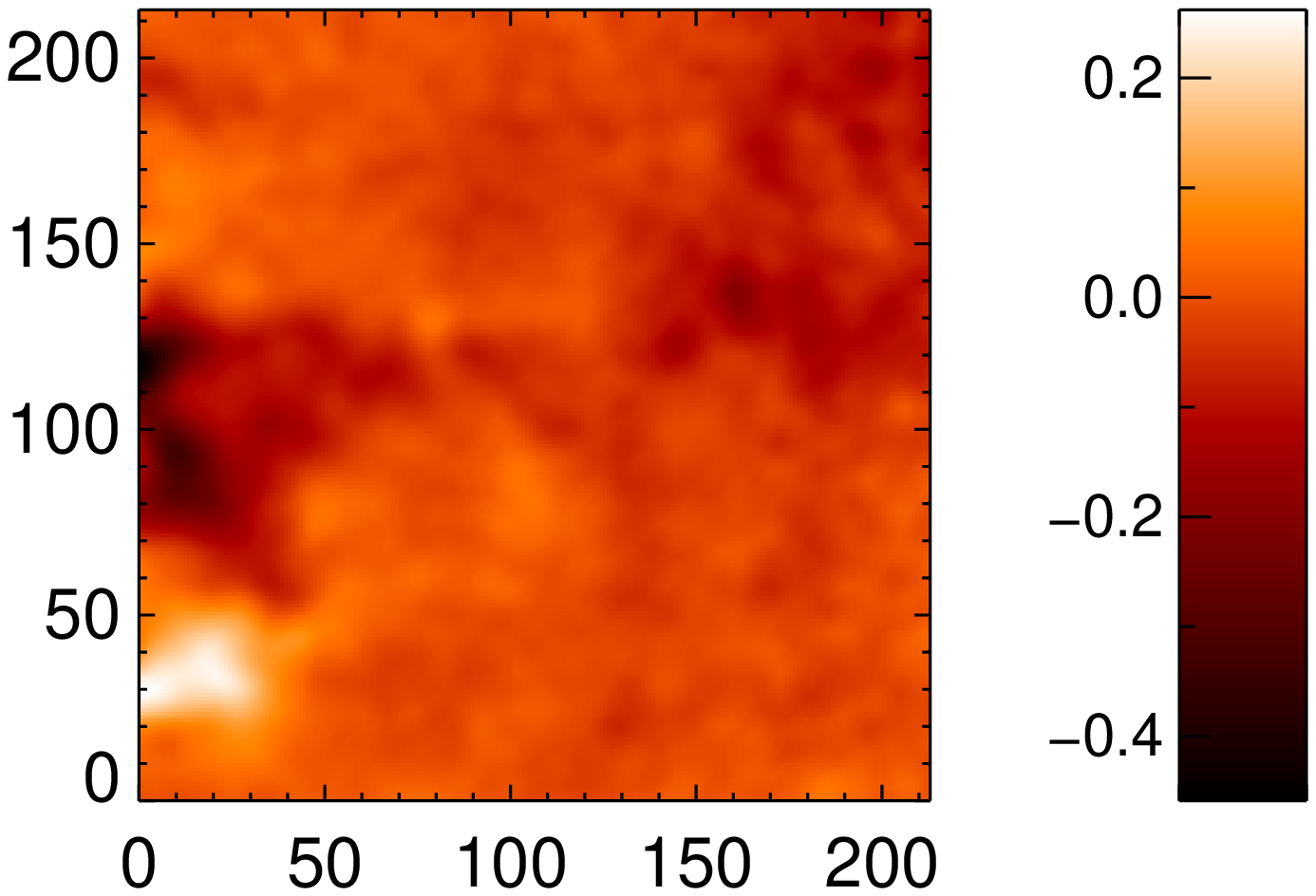} &
\includegraphics[width=4.8cm]{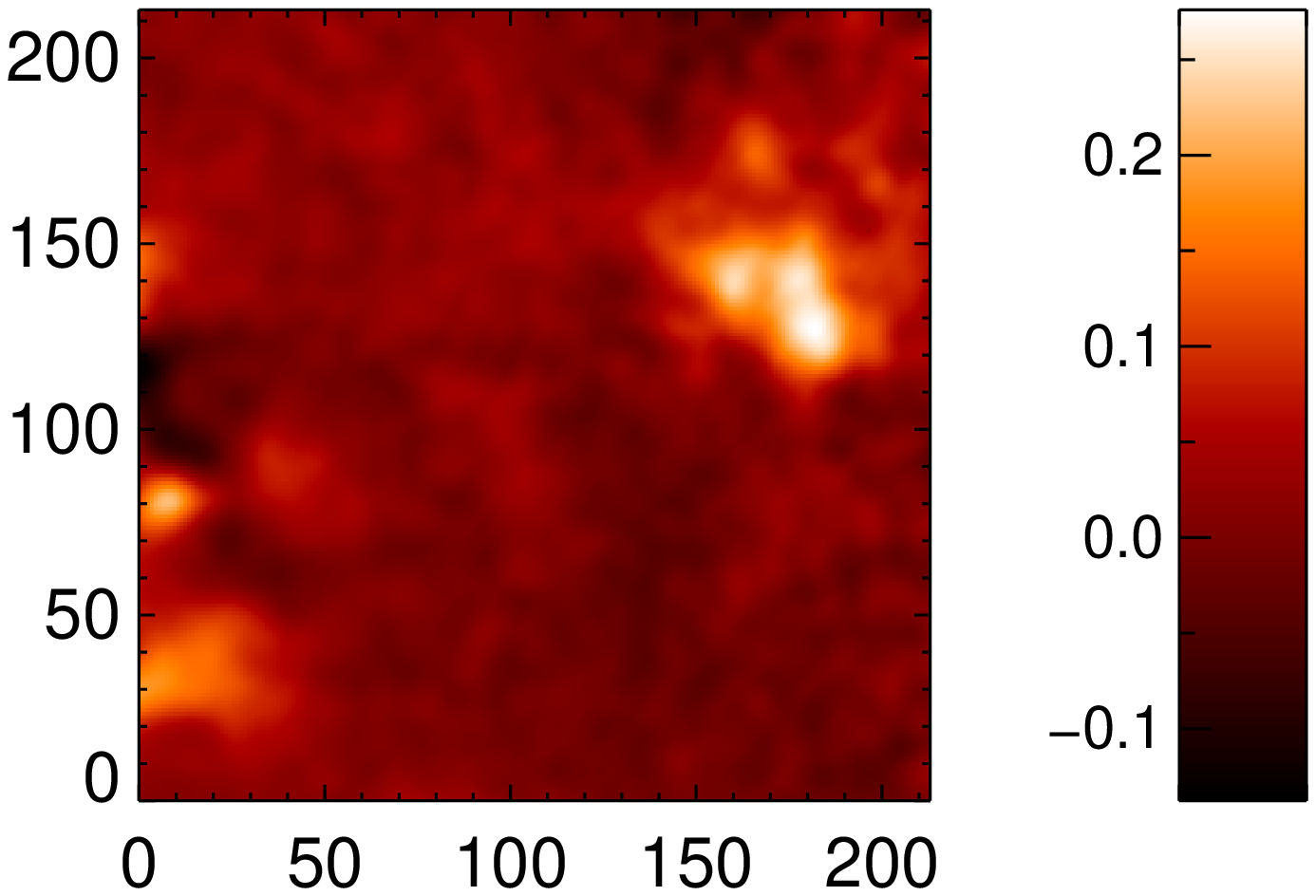} \\
\end{tabular}
\caption{\label{fig:eigenimages} Top row: The first, second and third eigenimages (left to right) for the $^{12}$CO data for Box 4. 
Middle row: as top row but for $^{13}$CO. Bottom row: as top row but for C$^{18}$O.}
\end{figure*}

The first three eigenimages for each of $^{12}$CO, $^{13}$CO and C$^{18}$O for Box 4 are given in Figure \ref{fig:eigenimages}. It can be seen that, comparing
with Figure 2 of \citet{brunt_etal_09}, the eigenimages here are typical of a turbulent cloud with large-scale driving. In order to investigate this more 
quantitatively the auto-correlation functions were calculated for each eigenimage and the typical auto-correlation lengths, $l_i$ where $i$ is the index of the 
eigenimage, were derived. As shown in \citet{brunt_etal_09}, the ratio of the correlation lengths of the first two eigenimages, $l_2/l_1$, is sensitive to the ratio of 
the turbulence driving length-scale, $\lambda_{\rm D}$ (say), to the cloud length-scale, $L_{\rm c}$. In our case the cloud length-scale should be taken to be the 
size of Box 4, or about 0.23\,pc. 

The sensitivity of $l_2/l_1$ to $\lambda_{\rm D}/L_{\rm c}$ is, perhaps unsurprisingly, lost once the driving scale becomes larger than the scale of the cloud.  
Generally speaking if $l_2/l_1 > 0.1$ -- 0.2 we can conclude that the turbulence is being driven at a scale larger than the cloud. Table \ref{table:numbers} contains these 
ratios for each of the data cubes, together with the range of values obtained when the analysis is performed for a total of 36 boxes, beginning with a box placed 10 pixels
South and 10 pixels East of the corresponding box shown in Fig.\ \ref{fig:pca-region} and incrementing the location by two pixels until the original box location is 
reached. This gives some indication of the sensitivity of the results to our placement of the boxes. Notwithstanding these ranges, it is clear that each of the ratios 
exceeds the above range by a considerable margin and thus we conclude that the driving scale of the turbulence here is larger than the size of our field. This is not an 
unusual finding \citep[e.g.][]{ossenkopf02}. Furthermore, assuming a sound speed of 0.25 -- 0.3 km s$^{-1}$ \citep{hartigan2022} the crossing time for Box 4,
the largest region analysed, is of order $10^6$\,yrs. Given that the size of Box 4 is a lower bound on the driving scale of the turbulence, this time-scale is 
a lower bound on the time over which we might expect un-driven turbulence to decay \citep[e.g.][]{ostriker01, downes09, downes11}. It is thus consistent to 
assume that the existing turbulence here is not driven primarily by radiative processes.

We can investigate this a little further by comparing the results we obtain from Boxes 1 and 2 with those from Box 3, since Box 3 is in Cloud B which is not irradiated
as strongly as the Western Wall. Our results suggest that for both \twco and \thco the ratio is lower in Cloud B than in the Western Wall, although not to such an extent
that we could conclude that the driving length-scale is of order the size of the cloud or less. However, \twco and \thco are both optically thick and so these only
probe the surface layers of the Western Wall and Cloud B. Thus we would expect that any effects of irradiation on the dynamics would be emphasised in these observations.
\ceio, on the other hand, is optically thin and here the results for Box 3 and Boxes 1 and 2 are consistent. We may conclude, then, that irradiation
is having an effect on the turbulence dynamics close to the surface of the Western Wall, but is not significantly impacting the interior of the cloud.

\begin{table}
\caption{\label{table:numbers} Ratio of auto-correlation lengths for the first and second eigenimages 
$(l_2/l_1)$ and the exponents for the first, second and third centroid velocity structure functions for each of $^{12}$CO, $^{13}$CO and C$^{18}$O. The subscripts on ``Box 4'' indicate the length-scales over which the fit to calculate the exponents was performed. For all
other regions the fits were performed over the range $[0.01,0.03]$\,pc. The range subscripts give the range of values for 
$(l_2/l_1)$ for the given box when the analysis is repeated for 36 boxes offset in two pixel increments from 10 pixels South and East of 
the boxes shown in Fig.\ \ref{fig:pca-region} up to the boxes shown.}
\centering
\begin{tabular}{@{}l|lllll}
\hline
Quantity & Region & $^{12}$CO & $^{13}$CO & C$^{18}$O \\ \hline
$(l_2/l_1)$ & Box 1 & 0.58 {\tiny [0.49,0.68]} & 0.94 {\tiny [0.86,1.03]} & 0.63 {\tiny [0.50,0.77]} \\
 & Box 2 & 1.14 {\tiny [0.46, 2.64]} & 0.51 {\tiny [0.47,0.56]} & 0.55 {\tiny [0.50,0.61]} \\
 & Box 3 & 0.7 {\tiny [0.36,1.13]} & 0.37 {\tiny [0.35,0.38]} & 0.51 {\tiny [0.44,0.58]} \\
 & Box 4 & 1.14 {\tiny [1.11,1.16]} & 0.86 {\tiny [0.80,0.90]} & 0.60 {\tiny [0.56,0.63]}\\ \hline
$\zeta_1$ & Box 1 & 0.63 & 0.54 & 0.47\\
          & Box 2 & 0.64 & 0.80 & 0.81 \\
          & Box 3 & 0.62 & 0.80 & 0.72 \\
          & Box 4 \tiny{($[0.01,0.03]$\,pc)} & 0.65 & 0.66 & 0.52 \\
          & Box 4 \tiny{($[0.03,0.055]$\,pc)}& 0.37 & 0.38 & 0.45 \\ \hline
$\zeta_2$ & Box 1 & 1.08 & 1.02 & 0.87 \\
          & Box 2 & 1.20 & 1.58 & 1.43 \\
          & Box 3 & 1.15 & 1.56 & 1.31 \\
          & Box 4 \tiny{($[0.01,0.03]$\,pc)} & 1.16 & 1.05 & 0.57 \\
          & Box 4 \tiny{($[0.03,0.055]$\,pc)}& 0.73 & 0.66 & 0.62 \\ \hline
$\zeta_3$ & Box 1 & 1.40 & 1.45 & 1.20 \\
          & Box 2 & 1.70 & 2.33 & 1.84 \\
          & Box 3 & 1.60 & 2.28 & 1.82 \\
          & Box 4 \tiny{($[0.01,0.03]$\,pc)} & 1.52 & 1.18  & 0.47 \\
          & Box 4 \tiny{($[0.03,0.055]$\,pc)}& 1.08 & 0.82 & 0.51 \\ \hline
$\beta$ & Box 1 & 2.08 & 2.02 & 1.87 \\
 & Box 2 & 2.20 & 2.58 & 2.43 \\ 
 & Box 3 & 2.15 & 2.56 & 2.31 \\ 
 & Box 4 \tiny{($[0.01,0.03]$\,pc)} & 2.16 & 2.18  & 1.47 \\
 & Box 4 \tiny{($[0.03,0.055]$\,pc)}& 2.08 & 1.82 & 1.51 \\ \hline
\hline
\end{tabular}

\medskip
$\beta$ is derived from the observed second order structure function as an approximation of the kinetic energy power spectrum, $E(k) \propto k^{-\beta}$. Note that this
approximation should be treated with caution since the observations do not contain the full 3D information on the velocity and density fields.
\end{table}

\cite{brunt_etal_09} suggest that values for $l_2/l_1$ far in excess of 0.2 might be accounted for by opacity
effects. The differences in the opacities of the \twco, \thco and \ceio, however, are not necessary to invoke in this case as the values observed for this ratio 
fall within the expected ranges from simulations.

%
%

\subsection{Structure functions}
\label{sec:structure-fn-results}
%
%
We first address the issue of the potential impact of large-scale velocity gradients on our calculations of the centroid velocity
increment structure function. We then consider the structure function scalings measured from our observations, compare these 
scalings with previous work and, finally, use the structure functions to deduce some properties of the turbulence from the 
observations.

\subsubsection{Effect of large-scale velocity gradient}

\begin{figure}
\centering
\includegraphics[width=8.4cm]{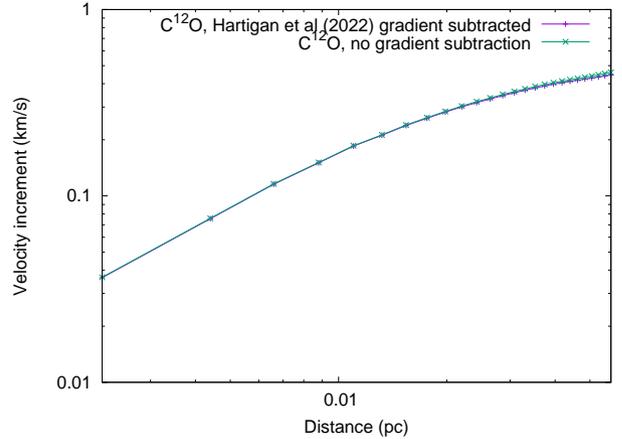} 
\caption{\label{fig:grad_twco} Plots of the first order structure function for $^{12}$CO for Box 4 in the case where the 2\kms
South to North gradient noted in Hartigan et al (2022) is subtracted, and without this subtraction.}
\end{figure}

\begin{figure*}
\centering
\begin{tabular}{cc}
\includegraphics[width=6.4cm]{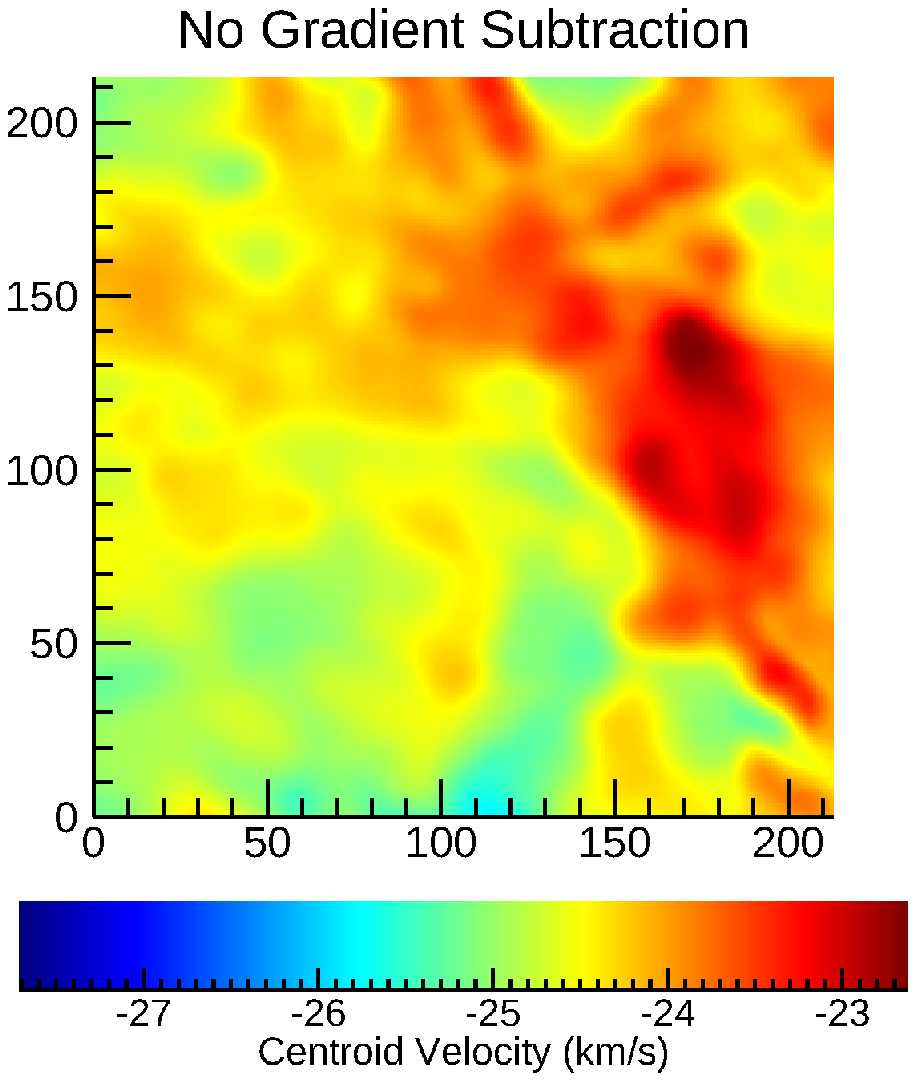} & \includegraphics[width=6.4cm]{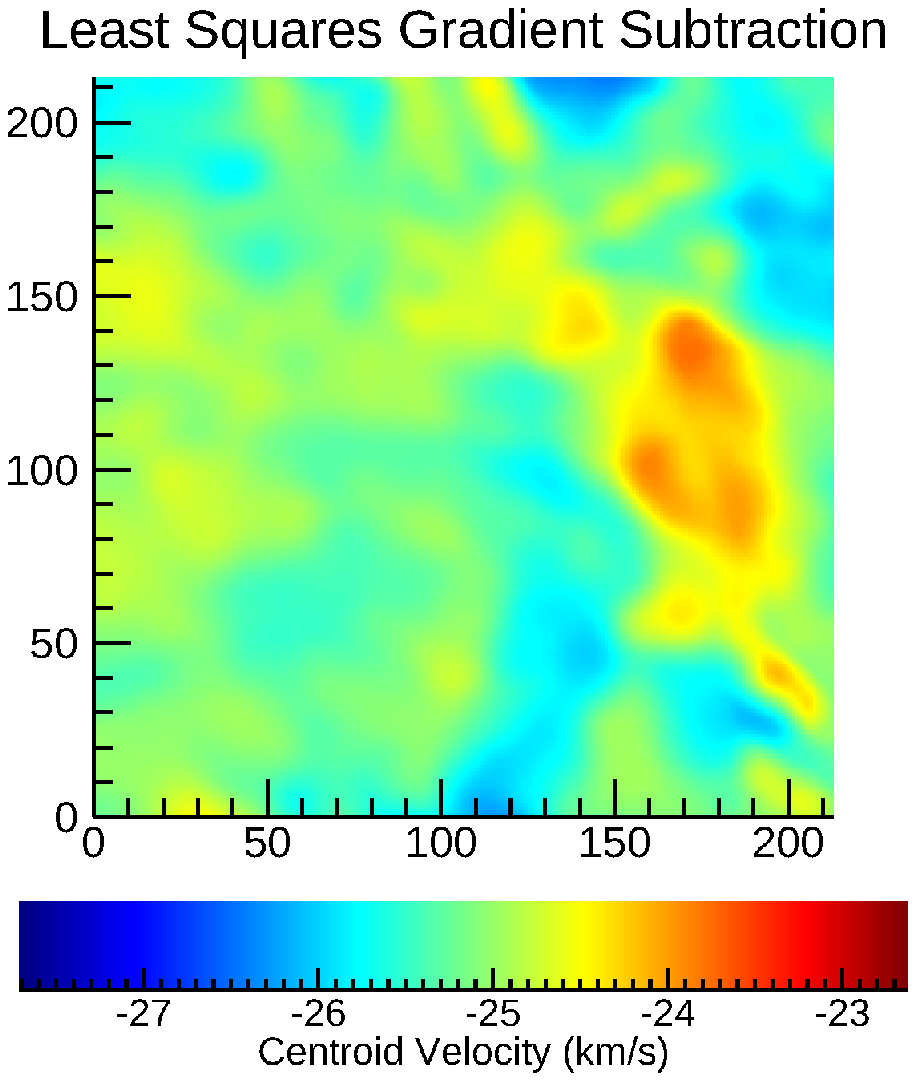} \\
\includegraphics[width=6.4cm]{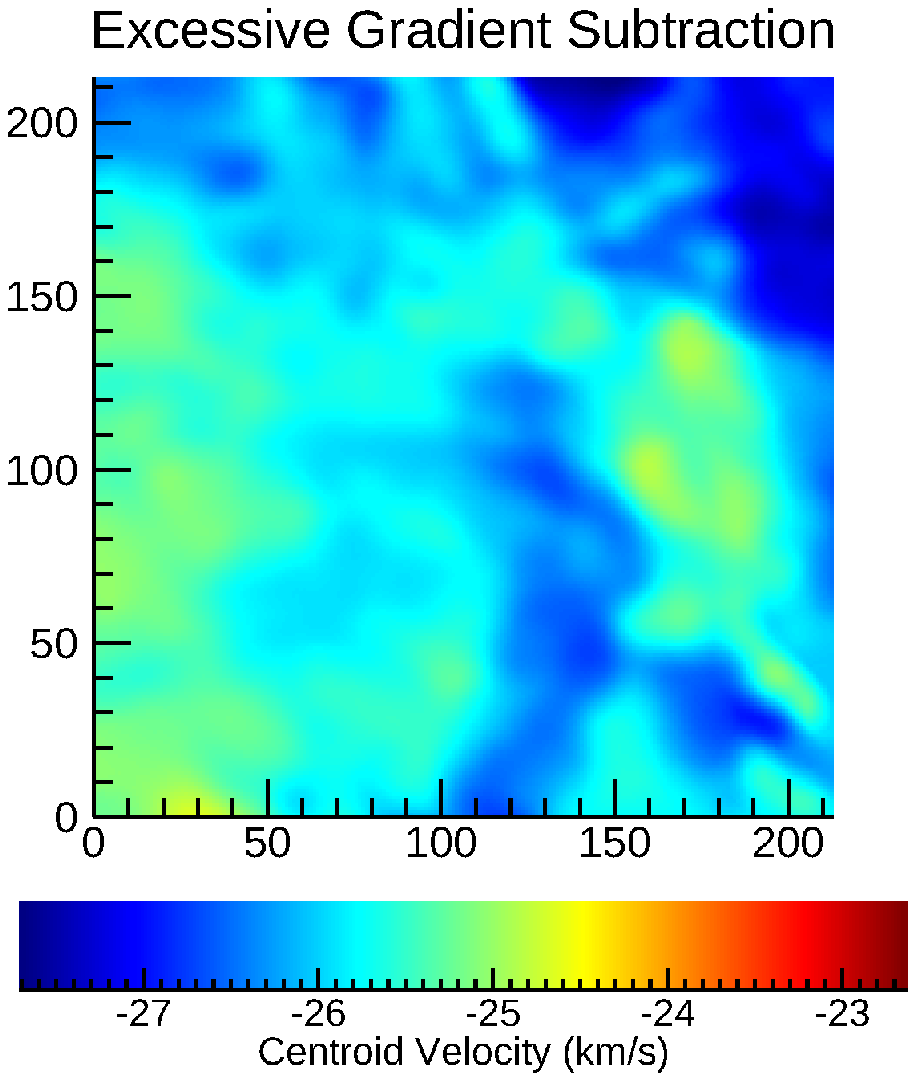} & \includegraphics[width=7.4cm]{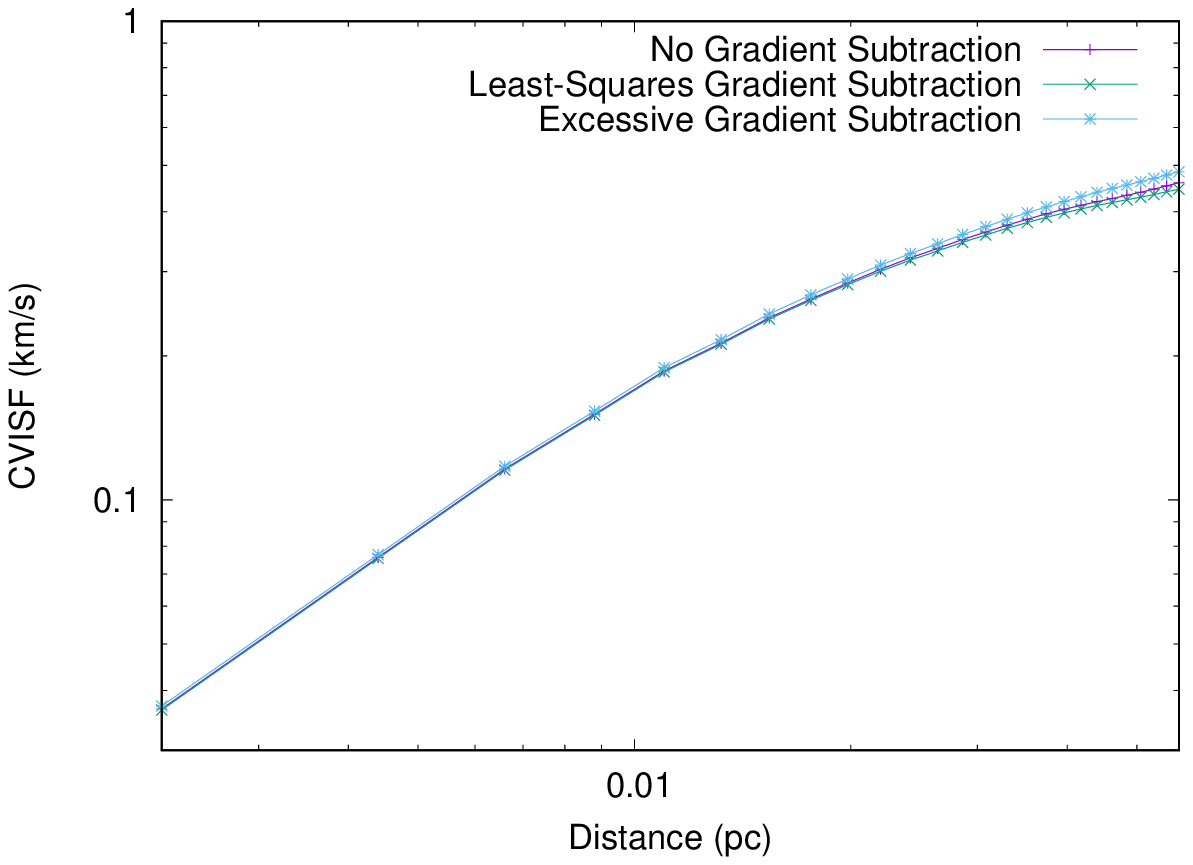} \\
\end{tabular}
\caption{\label{fig:grad_twco_lsq} Plots of the centroid velocity for \twco for Box 4 for the original centroid velocity (top left), 
the centroid velocity after subtraction of a least-squares fit plane (top right), the centroid velocity after subtraction of an 
excessive gradient (bottom left), and the centroid velocity increment structure functions for all three (bottom right). The gradient
subtractions make negligible difference to the structure functions (see text).}
\end{figure*}

If there is a large-scale velocity gradient in our data then this can impact the centroid velocity increment structure functions. If this
large-scale motion is ordered (e.g.\ solid-body rotation) then clearly it is not part of the turbulent cascade and should be removed from
our centroid velocity data prior to calculating the centroid velocity increment structure function \citep[e.g.][]{federrath16b}. Identifying
and subtracting cloud-scale gradients is not unusual and can lead to non-negligible corrections to the structure functions 
\citep[e.g.][]{menonetal_21, stewart_22}.

\cite{hartigan2022} note a 2\kms South to North gradient in the clump velocities across this entire observed field. We obviously wish to 
avoid over-estimating the turbulent velocities and, potentially, mis-characterising the turbulent cascade if the exponents of the 
structure functions are affected (see Sect.\ \ref{sec:structure_scaling}). The largest analysed field is Box 4 and is thus the
field which is most prone to such errors. The structure functions are calculated up to a length-scale of one quarter of this box and hence 
the largest length-scale in our structure functions is about 6\% of the overall extent of our field. We can estimate the expected impact of 
this on our centroid structure functions by doing the appropriate averaging analytically. This yields a maximum difference between the
gradient-subtracted and the original centroid velocity increment structure functions of around 0.06\,\kms, less than the velocity resolution 
of the underlying data. Thus we do not expect this large-scale gradient to impact our results and figure \ref{fig:grad_twco}, which 
contains plots of the first order centroid velocity increment structure functions for \twco with and without this gradient subtraction
performed confirms this.

Since our observations do not include the full extent of the molecular cloud in the East-West direction we might be concerned that there may
be an ordered centroid velocity gradient in this direction which is not obvious in our data. To test the possible effect of such a gradient 
we fitted a plane using least-squares to our \twco centroid velocity data and used this to investigate the impact of any such effect. Note
that this is intended only as an investigation: it requires considerable physical justification to remove such relatively small scale 
gradients by fitting a plane (as opposed to any other type of surface), and without knowing whether they are associated with a turbulent 
cascade. In any case, this had negligible impact on our data, so we then performed an excessive subtraction in which we, arbitrarily, 
hypothesise that a gradient twice that of the least-squares estimate is present.  Figure \ref{fig:grad_twco_lsq} contains images of the 
centroid velocity in \twco for Box 4, the gradient-subtracted centroid velocity, and the excessive gradient subtracted centroid velocity, 
and plots of the centroid velocity increment structure function for each. The exponents calculated for each of these structure functions 
(see next section) differed by less than 2\%. Thus velocity gradients have negligible impact in our data. This is not unusual and has been 
found previously in, for example, the analysis of Pillar 20 in \citet{menonetal_21}.

\subsubsection{Structure function scaling}
\label{sec:structure_scaling}

\begin{figure*}
\centering
\begin{tabular}{cccc}
\includegraphics[width=5.0cm]{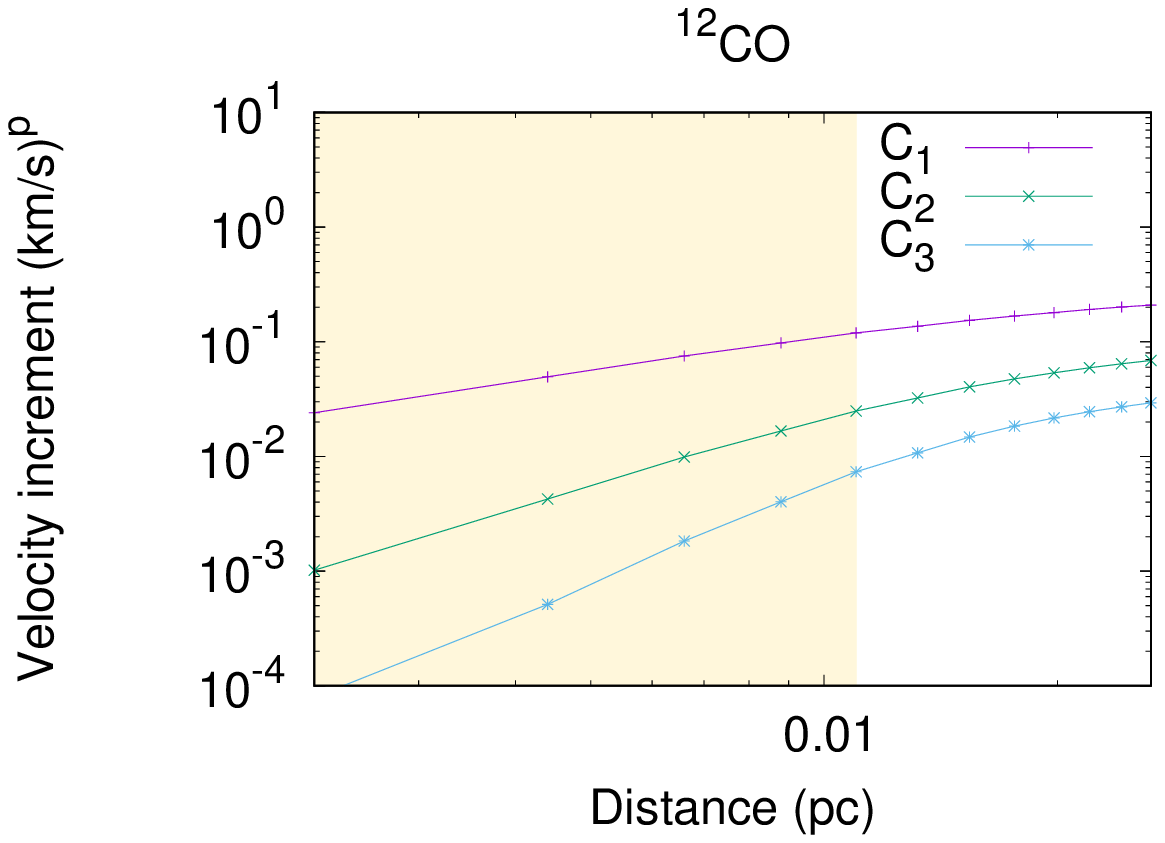} & \includegraphics[width=5.0cm]{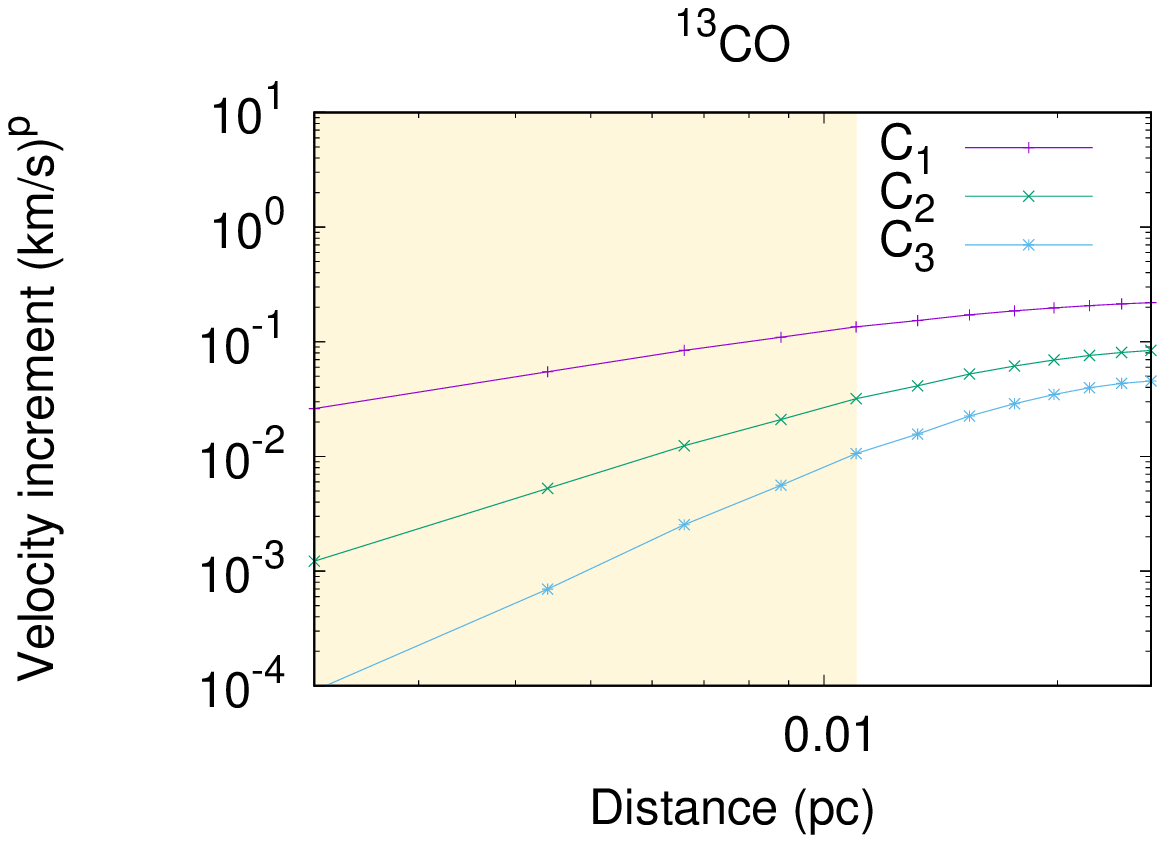} &
\includegraphics[width=5.0cm]{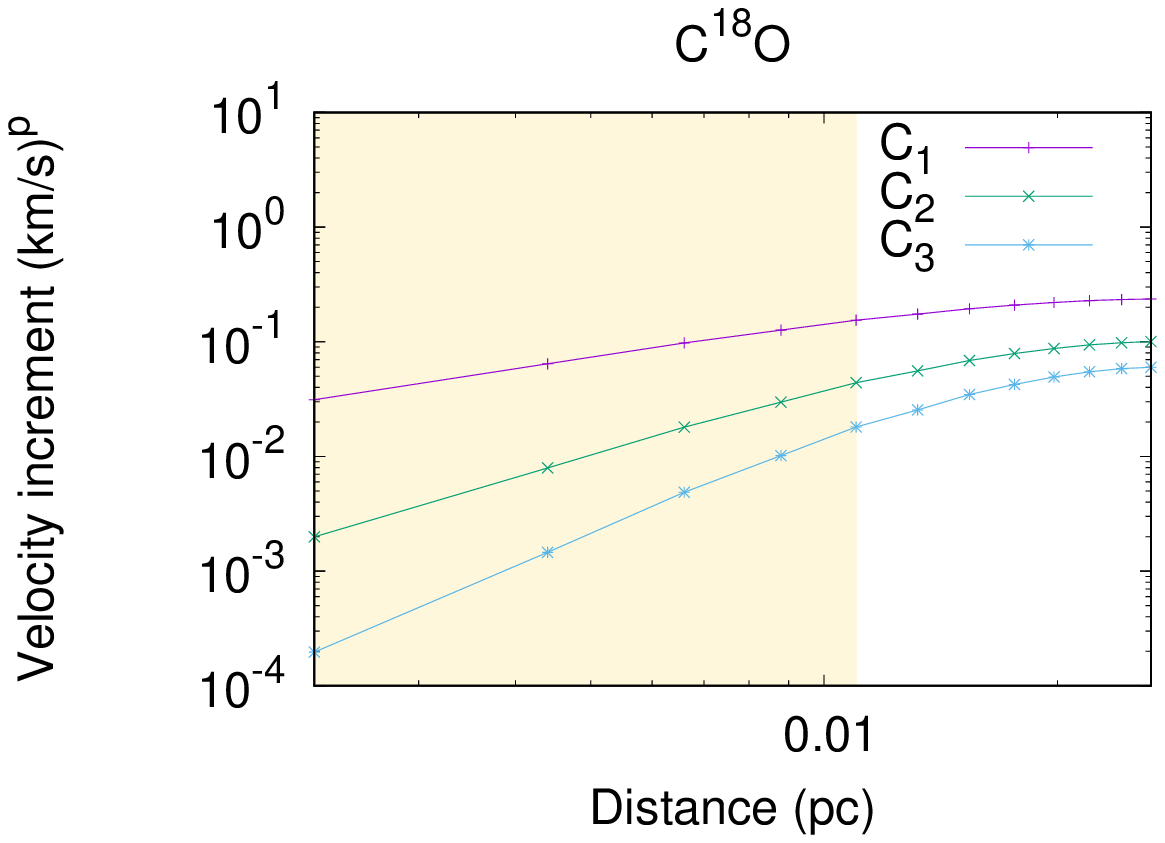} \\
\includegraphics[width=5.0cm]{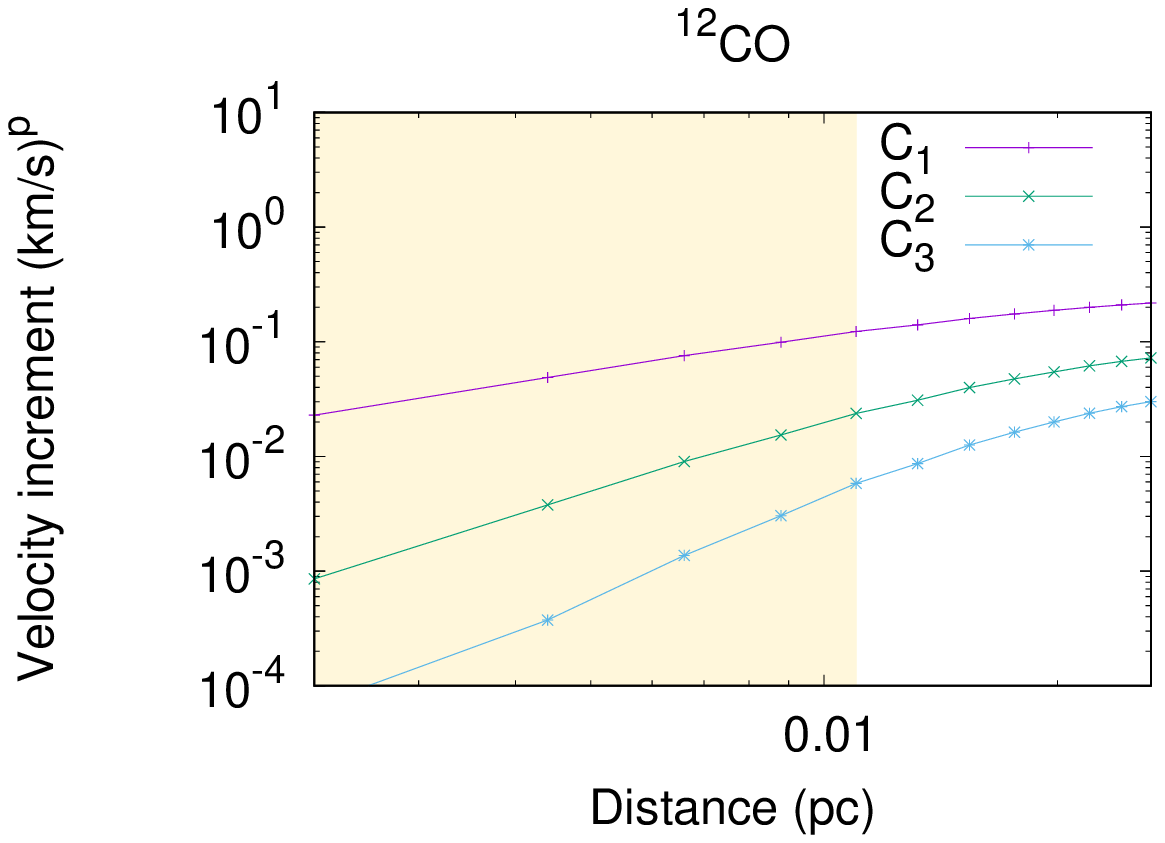} & \includegraphics[width=5.0cm]{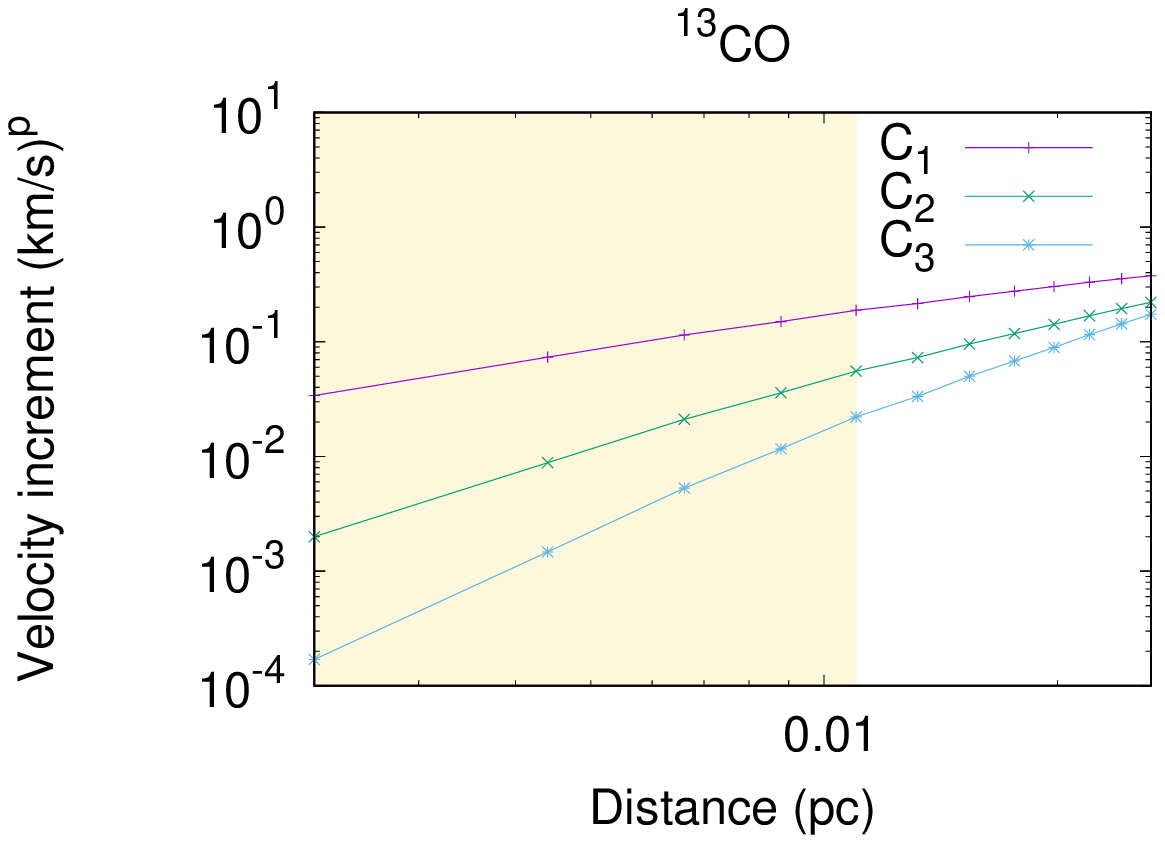} &
\includegraphics[width=5.0cm]{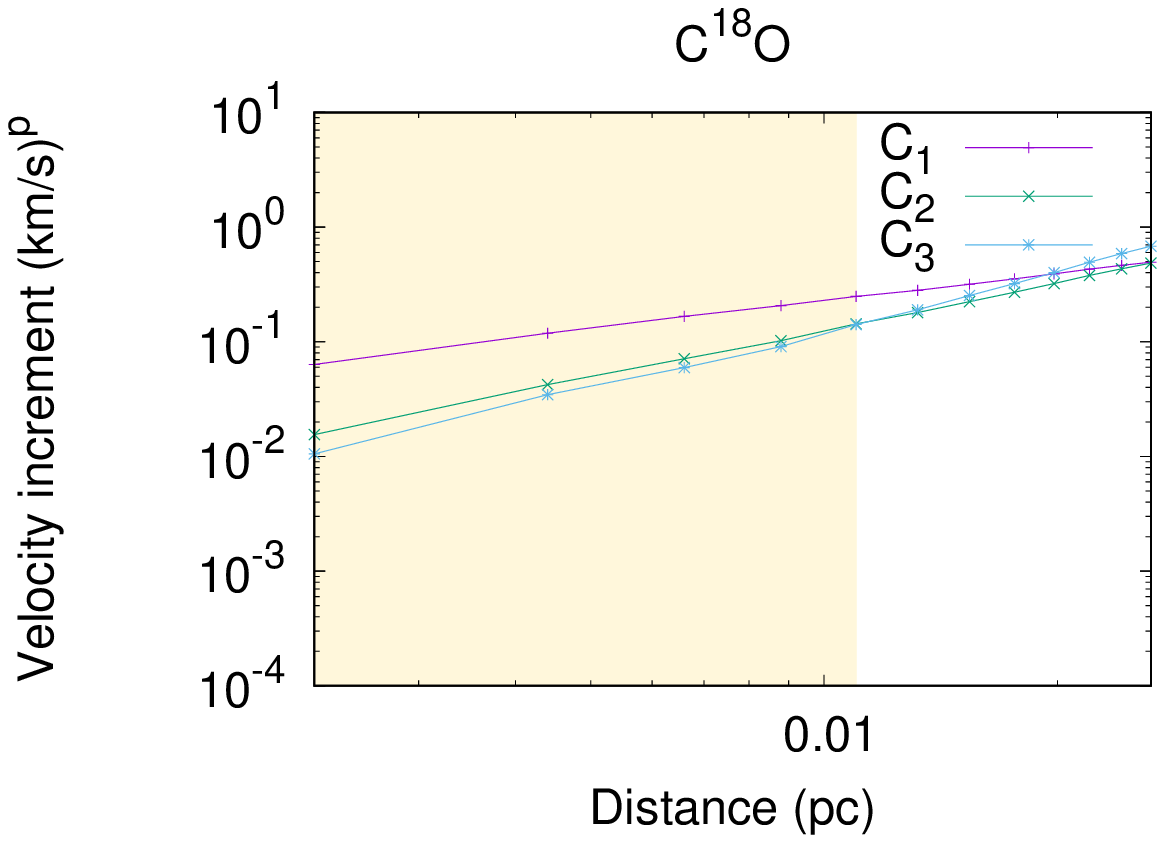} \\
\includegraphics[width=5.0cm]{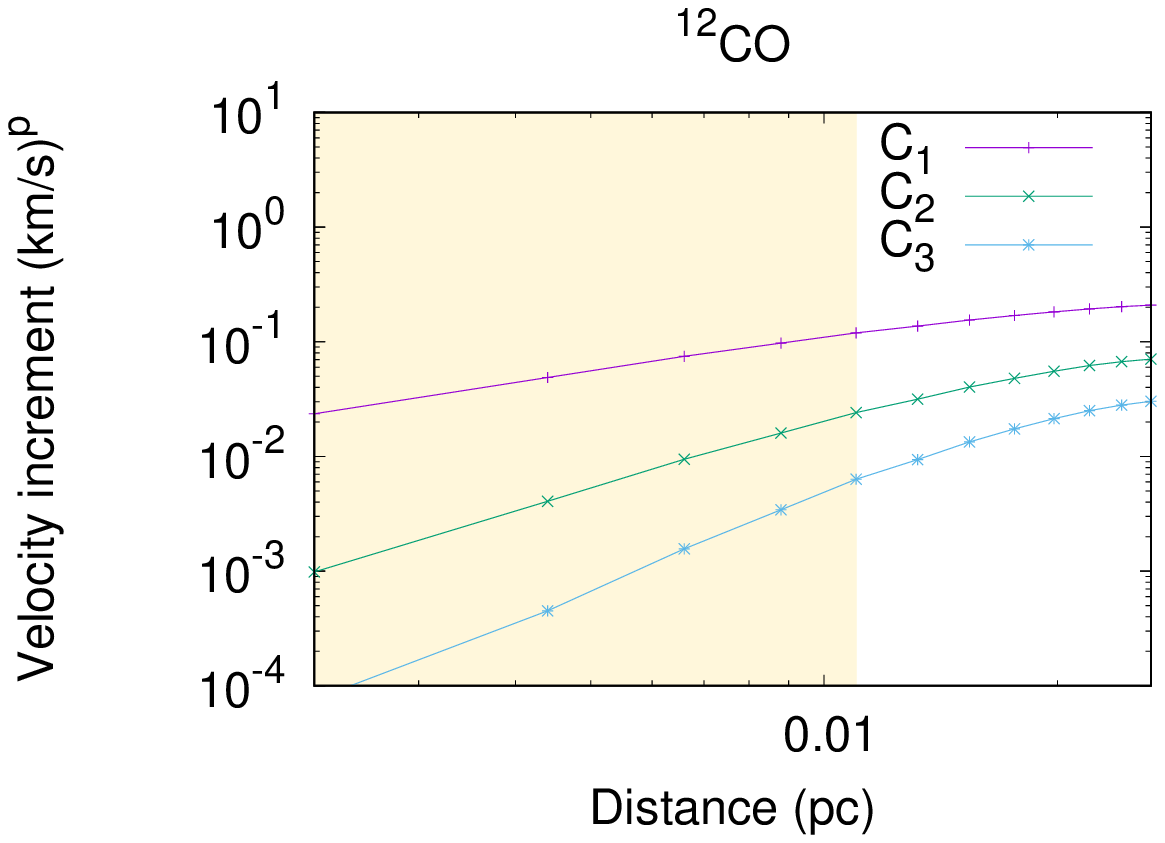} & \includegraphics[width=5.0cm]{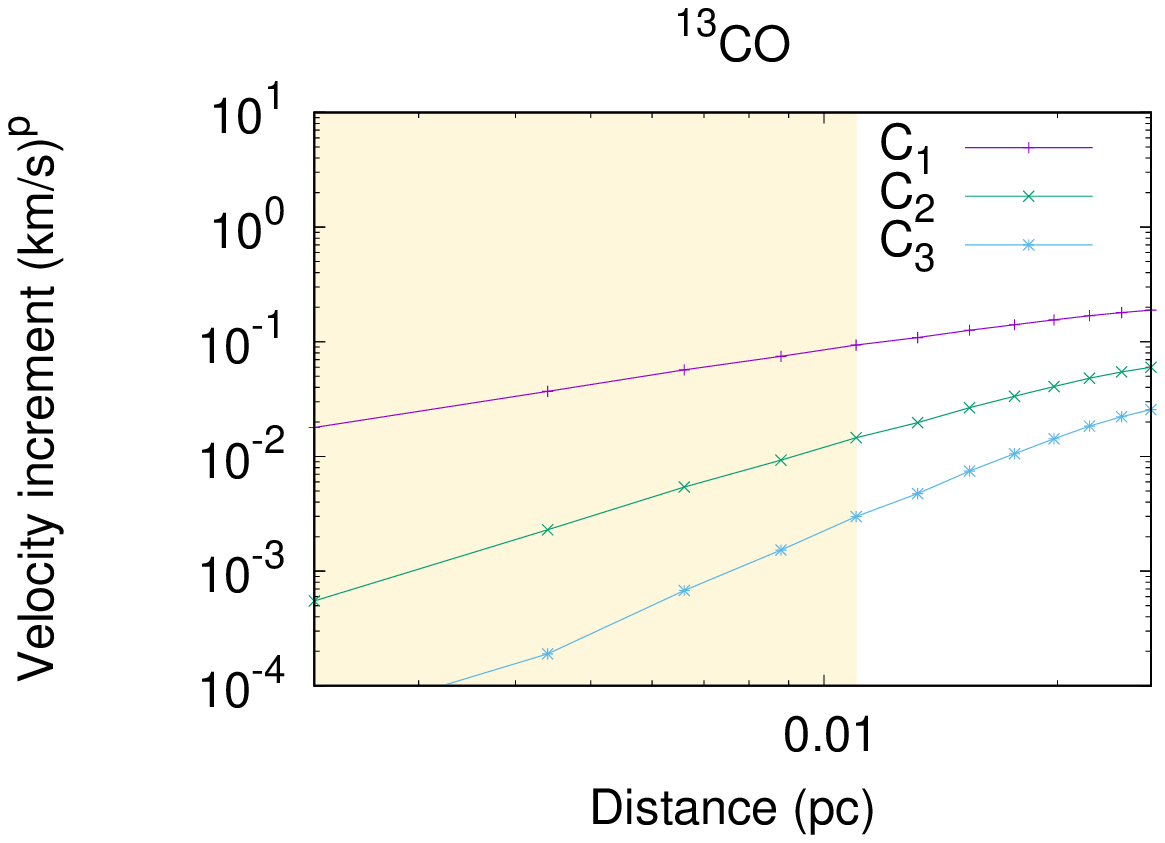} &
\includegraphics[width=5.0cm]{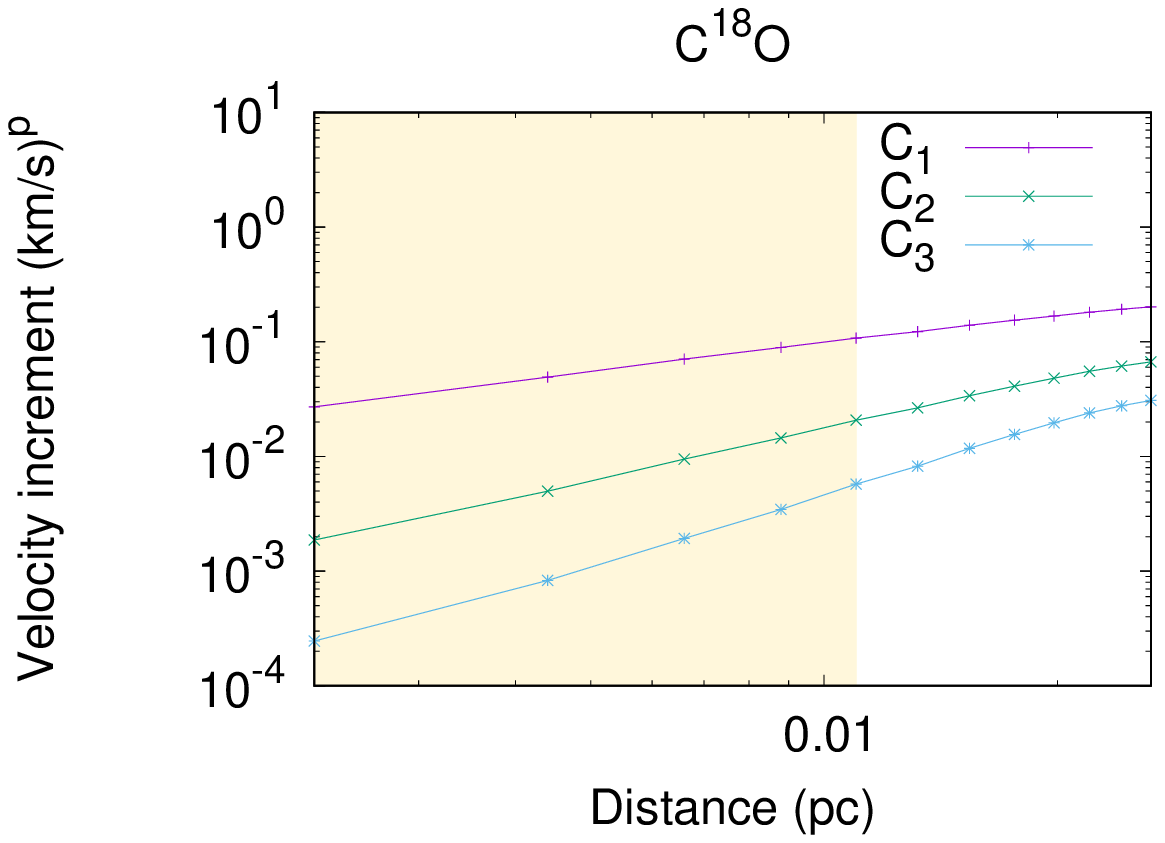} \\
\includegraphics[width=5.0cm]{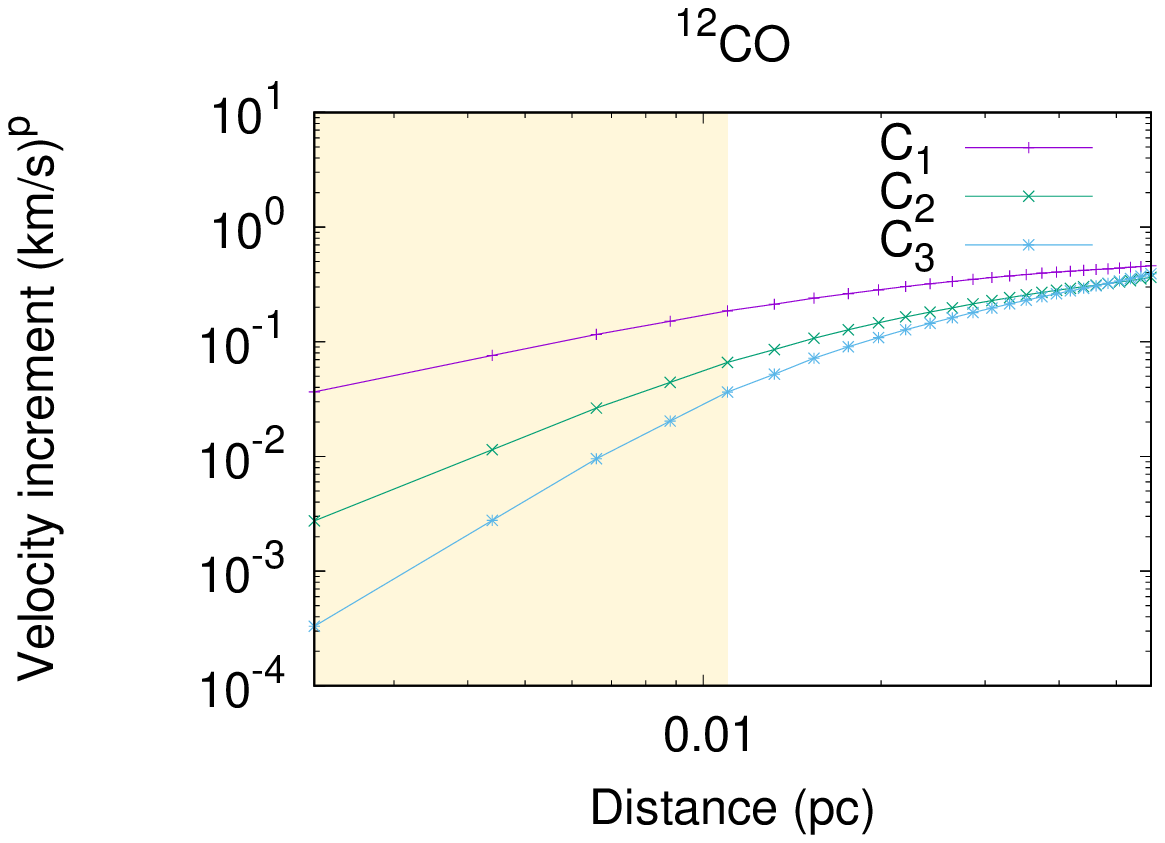} & \includegraphics[width=5.0cm]{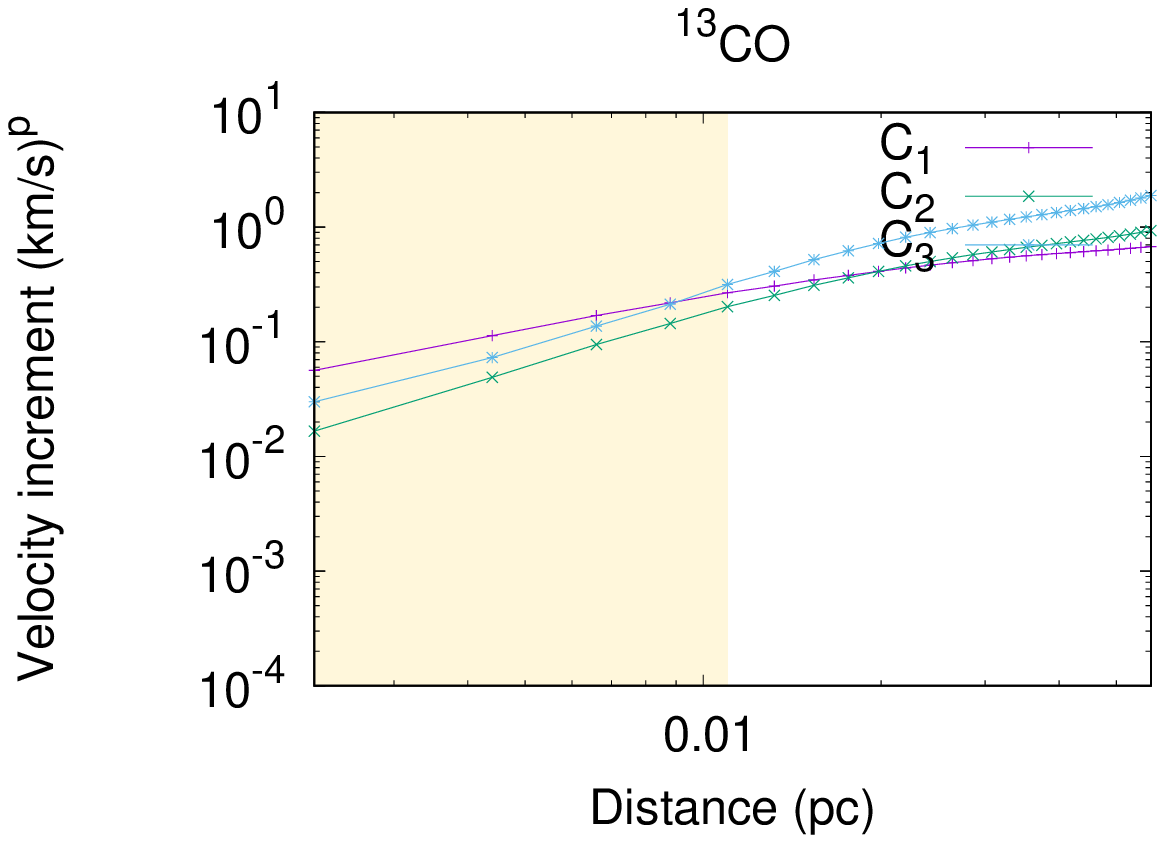} &
\includegraphics[width=5.0cm]{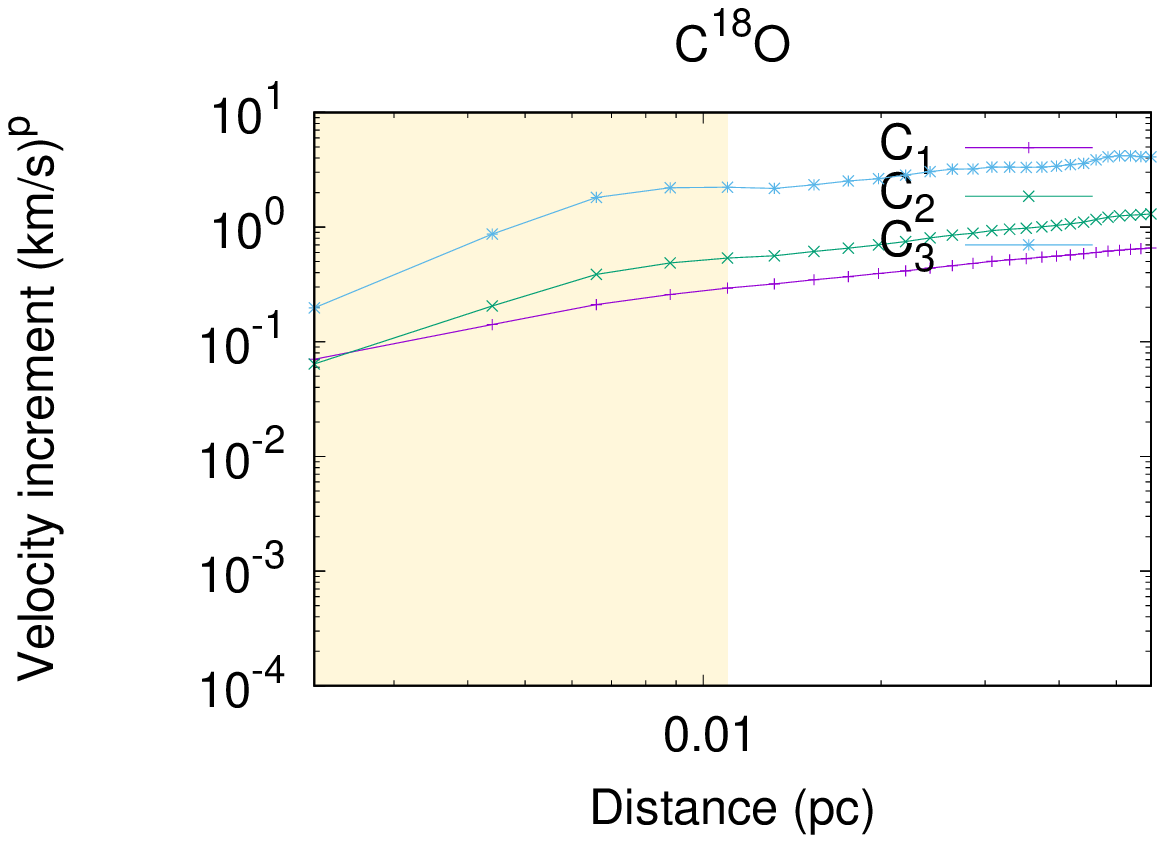}
\end{tabular}
\caption{\label{fig:structure_fns} Plots of the first, second and third order centroid velocity structure 
functions for Boxes 1-- 4 (top row to bottom row, respectively) for each of $^{12}$CO, $^{13}$CO and C$^{18}$O (left to right). Note the larger range of 
length-scales in the plots for Box 4. The shaded region in each plot indicates that which is unresolved by the observations. The superscript $p$ in the 
axis label indicates the order of the structure function being plotted.}
\end{figure*}

Plots of the first three centroid velocity structure functions for Box 4 are shown in Figure \ref{fig:structure_fns} for our
$^{12}$CO, $^{13}$CO and C$^{18}$O data. We expect that these functions should be power laws in distance 
\citep[e.g.][]{frisch95, boldyrev02, brunt_etal03}: 
\begin{equation}
C_l(r) \propto r^{\zeta_l},
\end{equation}
where $l$ is the order of the structure function. While it is possible to convince ourselves that Fig.\ 
\ref{fig:structure_fns} demonstrates this, at least in the resolved region, a close examination of the second and third order structure
functions for the $^{12}$CO and \thco structure functions suggests a possible break in the power-law between 0.02\,pc and 0.03\,pc. This length is of 
interest, as we shall see in Sect.\ \ref{sec:emission-length-scales}. However, the \ceio structure functions show no such break.  
Table \ref{table:numbers} contains the exponents of the first, second and third order structure functions for each of Boxes 1, 2 and 3 under the assumption that they are 
power laws fitted over the range $[0.01,0.03]$\,pc. For Box 4 these exponents are derived by fitting power laws separately over the ranges $[0.01,0.03]$\,pc and 
$[0.03,0.055]$\,pc. In the range $[0.01,0.03]$\,pc the \twco and \thco data have similar exponents to each other, while the \ceio exponents differ significantly, at least 
for the second and third order functions.

%
%

\begin{figure}
\centering
\includegraphics[width=8.4cm]{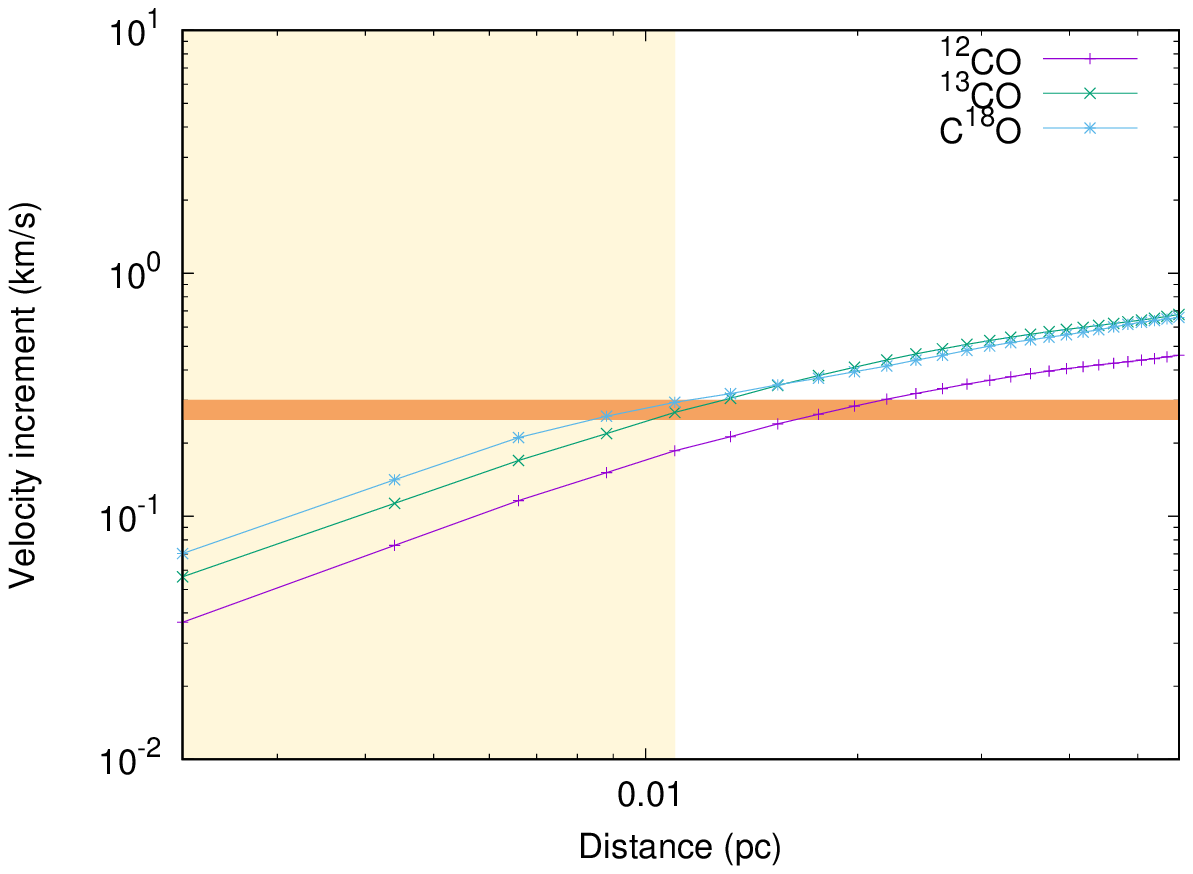} 
\includegraphics[width=8.4cm]{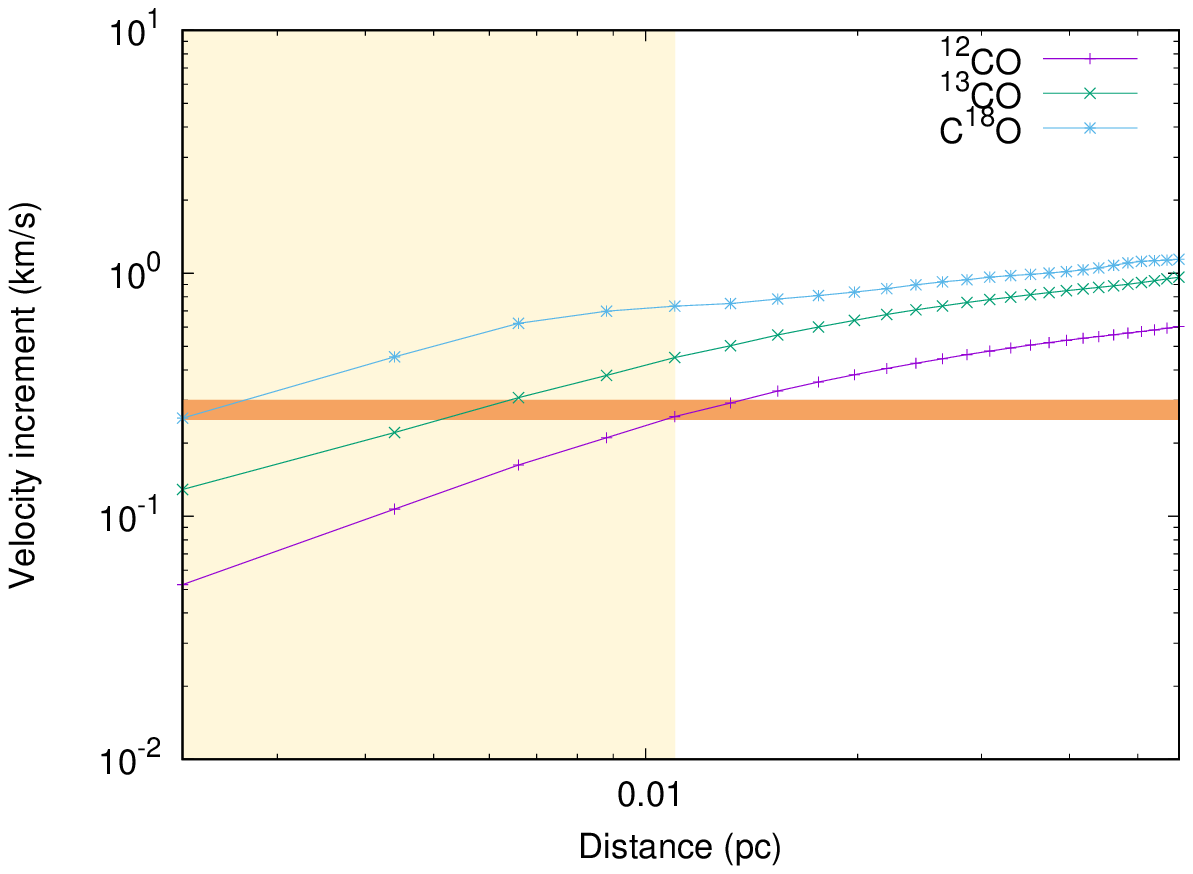}
\caption{\label{fig:structure_fns_scales} Plots of the first (top panel) and square root of the second (bottom panel) structure functions 
for each of $^{12}$CO, $^{13}$CO and C$^{18}$O data for Box 4. The shaded region 
to the left of each plot indicates the length-scales which are unresolved by the observations, while the horizontal
shaded region indicates the velocity range within which the sound speed lies.}
\end{figure}

In order to ease comparison between the different isotopes, figure \ref{fig:structure_fns_scales} contains 
plots of the first order, and square-root of the second order, structure functions for Box 4. The likely range of the sound speed is 
also indicated on these plots. The characteristic centroid velocities at each length implied by the structure functions for the different isotopes are 
somewhat different.  However, in all cases the sonic scale implied for the turbulence is between 0.01 and 0.02\,pc for each isotope (for the 
first order structure function).  

The velocity resolution is approximately 0.166\,\kms which is close to the sound-speed. Hence our estimated range 0.01 -- 0.02\,pc for the
sonic scale should be regarded as an upper limit. \citet{federrath16a} found typical sonic scales of around 0.04 -- 0.16\,pc, up to an order of
magnitude larger than the upper limits inferred here.  We come back to this point in Sect.\ \ref{sec:emission-length-scales}.

There are notable differences between the C$^{18}$O structure functions and those of $^{12}$CO and 
$^{13}$CO. Firstly it is clear that a striking flattening of the third order C$^{18}$O structure function occurs at
distances even shorter than the resolution of the data, and this does not occur for the other CO isotopes. 
Flattening of this type often occurs when the distances involved are of order the driving scale of the turbulence. If we
consider the first and second order structure functions, it is again clear that the \ceio functions increase much less with
distance.

There are a variety of potential explanations for this result. There is a difference in opacity 
between the emission for the different isotopes \citep{hartigan2022}.  Since \ceio emission suffers less absorption 
it is likely to suffer more from projection smoothing, whereby small scale motions along the line of sight are 
averaged away. In principle this could lead to the lower velocities in the \ceio structure functions when 
compared to the structure functions for the higher opacity \twco and \thco emission. The issue with this 
interpretation is, however, that in molecular cloud turbulence small scale motions are at relatively low 
velocities while the large scale motions, which do not suffer the same degree of projection smoothing, 
contribute primarily to the large velocities. Thus projection smoothing should preferentially effect the low velocity
parts of the structure functions, and we do not see evidence of this.

\begin{figure}
\centering
\includegraphics[width=8.4cm]{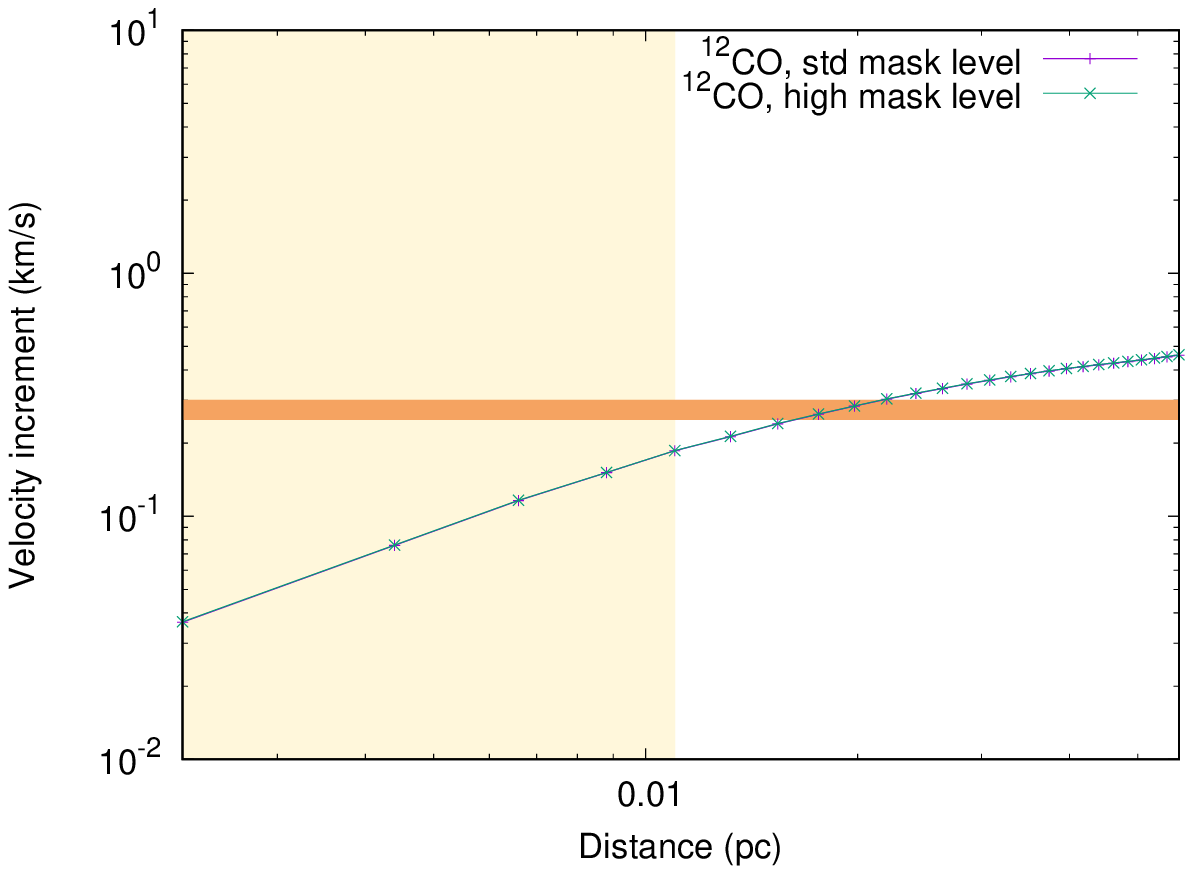} 
\includegraphics[width=8.4cm]{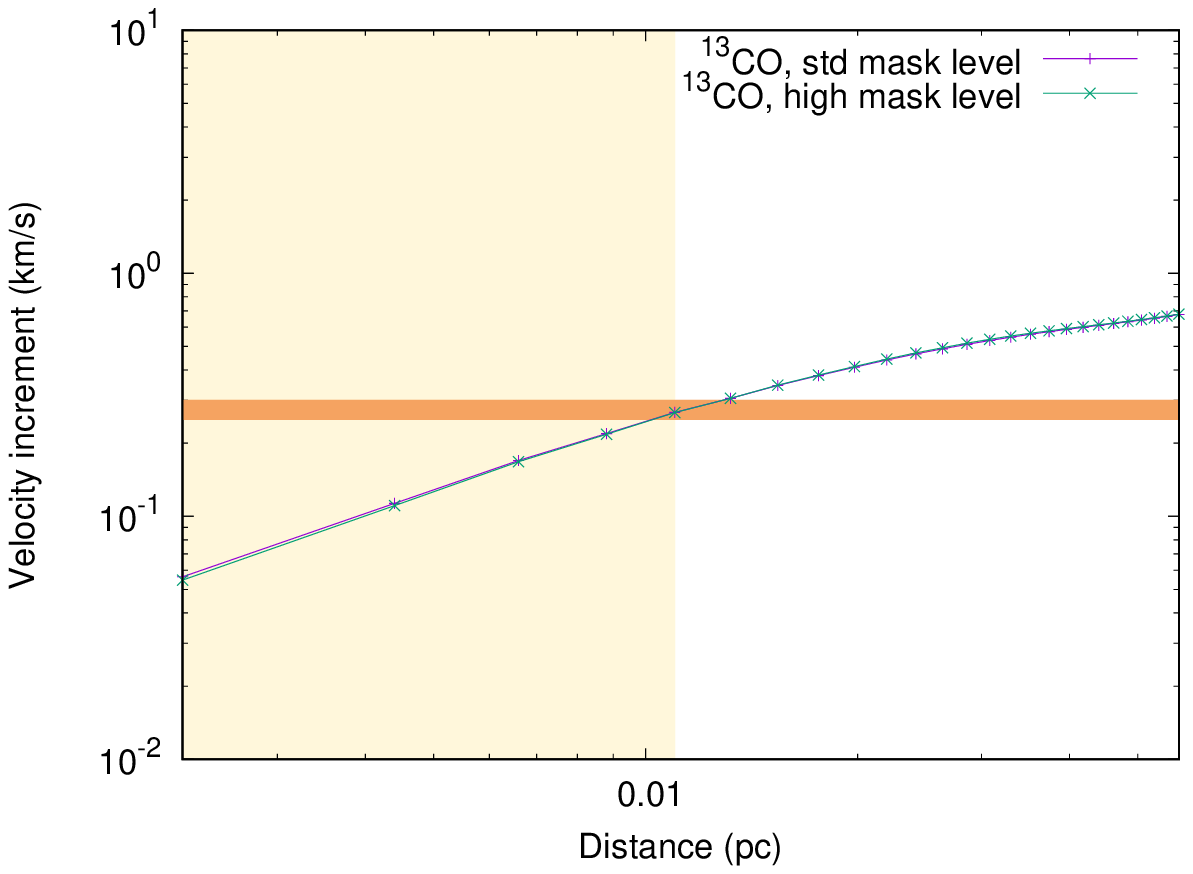} 
\includegraphics[width=8.4cm]{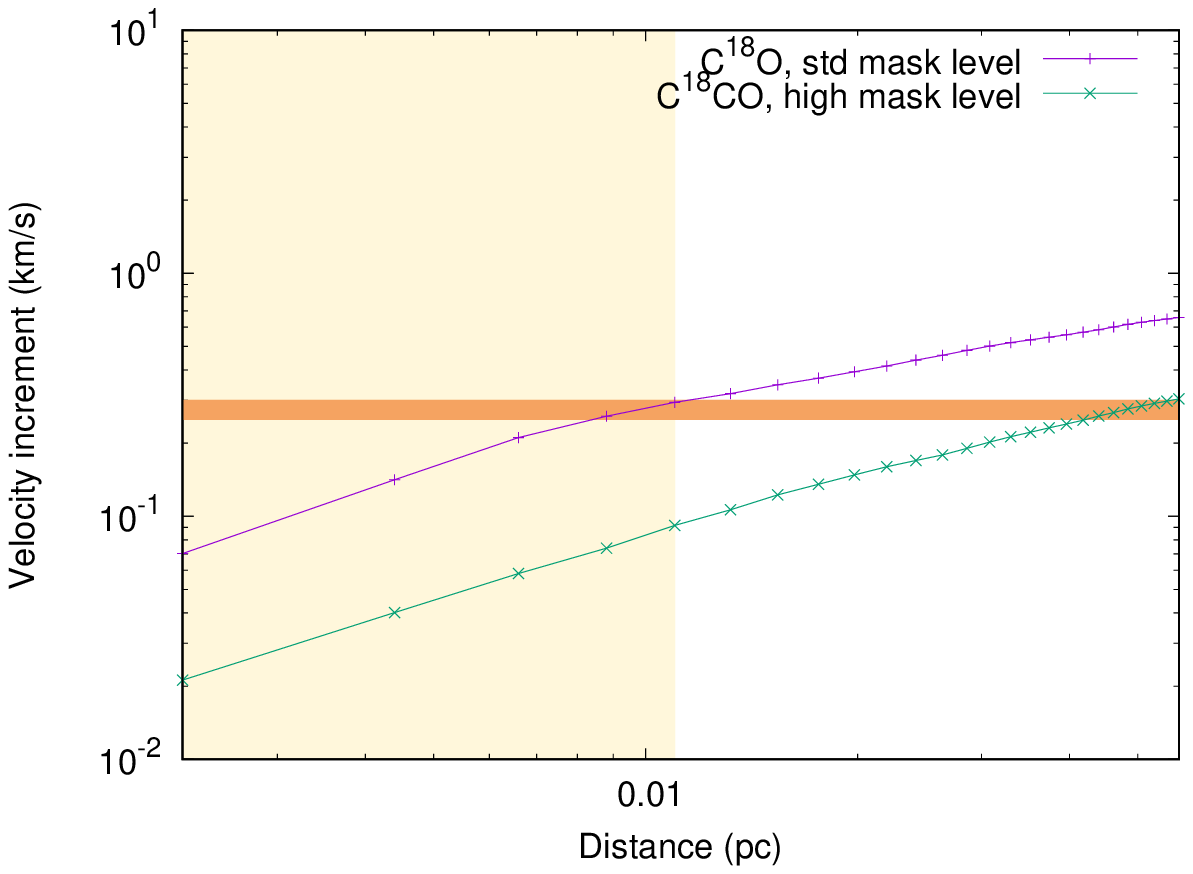} 
\caption{\label{fig:mask-level-comp} Plots of the first order centroid velocity structure 
functions for Box 4 for each of $^{12}$CO, $^{13}$CO and C$^{18}$O for the standard masking of five times the noise level in
the data, and for a high masking level of ten times the noise level (top to bottom). The shaded region in each plot 
indicates that which is unresolved by the observations.}
\end{figure}

A much more likely explanation becomes evident when we look at the spatial distribution of the emission from \ceio and 
compare this with the distributions for \twco and \thco (Fig.\ \ref{fig:pca-region}). It is clear that the 
\ceio emission is more highly localised towards the PDR and this means that, certainly at length-scales larger than this 
localised emission, and potentially at all length-scales, the \ceio structure function will be more highly influenced by 
noise. To test the likely impact of this in Figure \ref{fig:mask-level-comp} we plot the first order structure functions 
for Box 4 for all isotopes, and also for a higher masking level of 10 times the noise level, rather than 5 times this noise 
level (see Sec.\ \ref{sec:analysis}). It is clear that the masking has a very significant impact on the \ceio structure 
functions. The equivalent plots for \twco and \thco show essentially no impact of the masking level on the structure 
functions. We suggest that this is the most likely cause of the differences in the centroid velocities and, despite the 
opacity of the \twco and \thco lines, we speculate that these yield somewhat more reliable indicators of the underlying 
centroid velocity structure function than the \ceio emission.

\subsubsection{Comparison of structure function scaling with previous work}

Both \citet{federrathetal_10} and \cite{downes12} contain the power-law exponents of structure functions calculated from
hydrodynamic and multi-fluid magnetohydrodynamic simulations respectively. In addition, \citet{hily-blantetal_08} present
observed structure function exponents. Comparing the figures in Table \ref{table:numbers} with those of Table 4 in \citet{federrathetal_10}
and Table 5 in \cite{downes12}, and keeping in mind that our exponents are for a projected velocity structure function, we note that
our absolute scaling exponents lie between those of \citet{federrathetal_10} and \cite{downes12}. The scaling exponents normalised to the
exponent of the third order structure function \citep[see][for details]{federrathetal_10} are all rather large when compared with either set of results,
with the exception of the exponents calculated for the intermittency model of \cite{boldyrev02}. However, the spread of values calculated here is
consistent with both papers, and thus with simulations of hydrodynamic turbulence or multi-fluid magnetohydrodynamic turbulence.

\begin{figure*}
\centering
\begin{tabular}{cc}
\includegraphics[width=6.4cm]{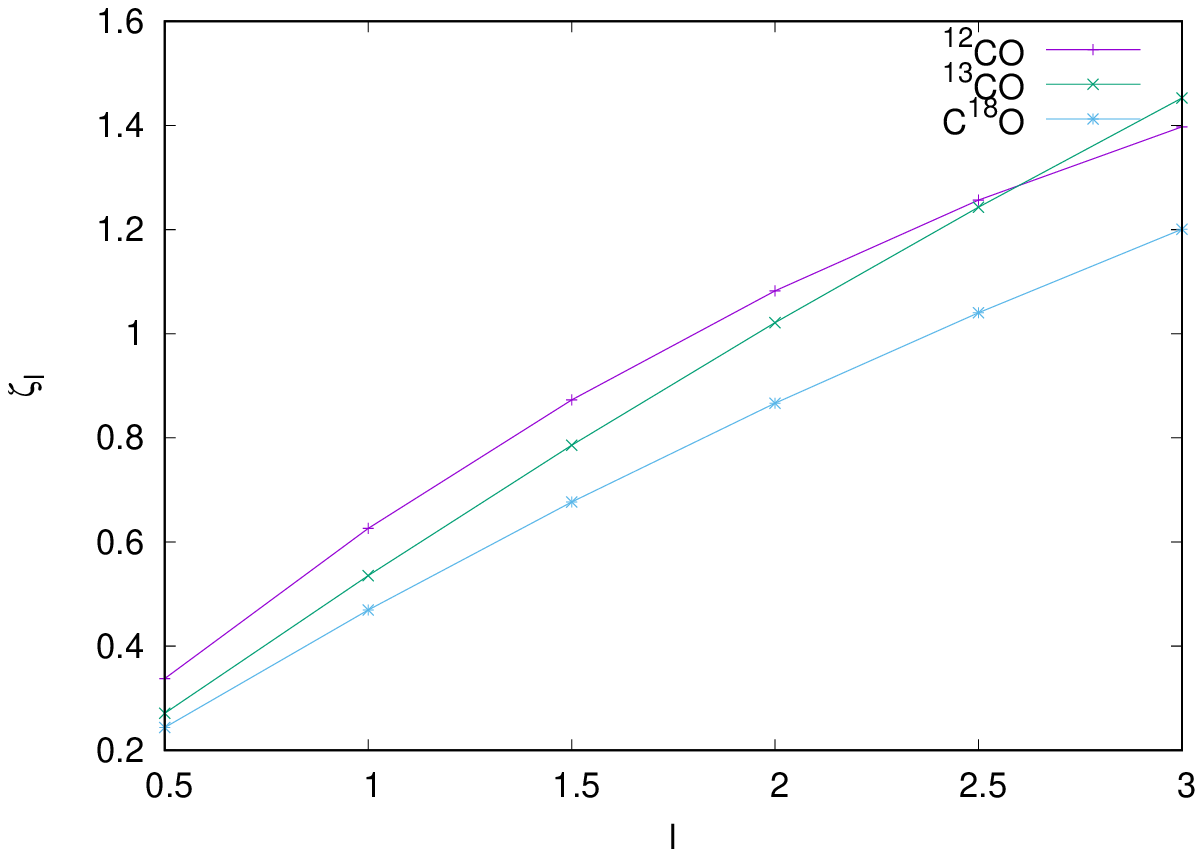} & \includegraphics[width=6.4cm]{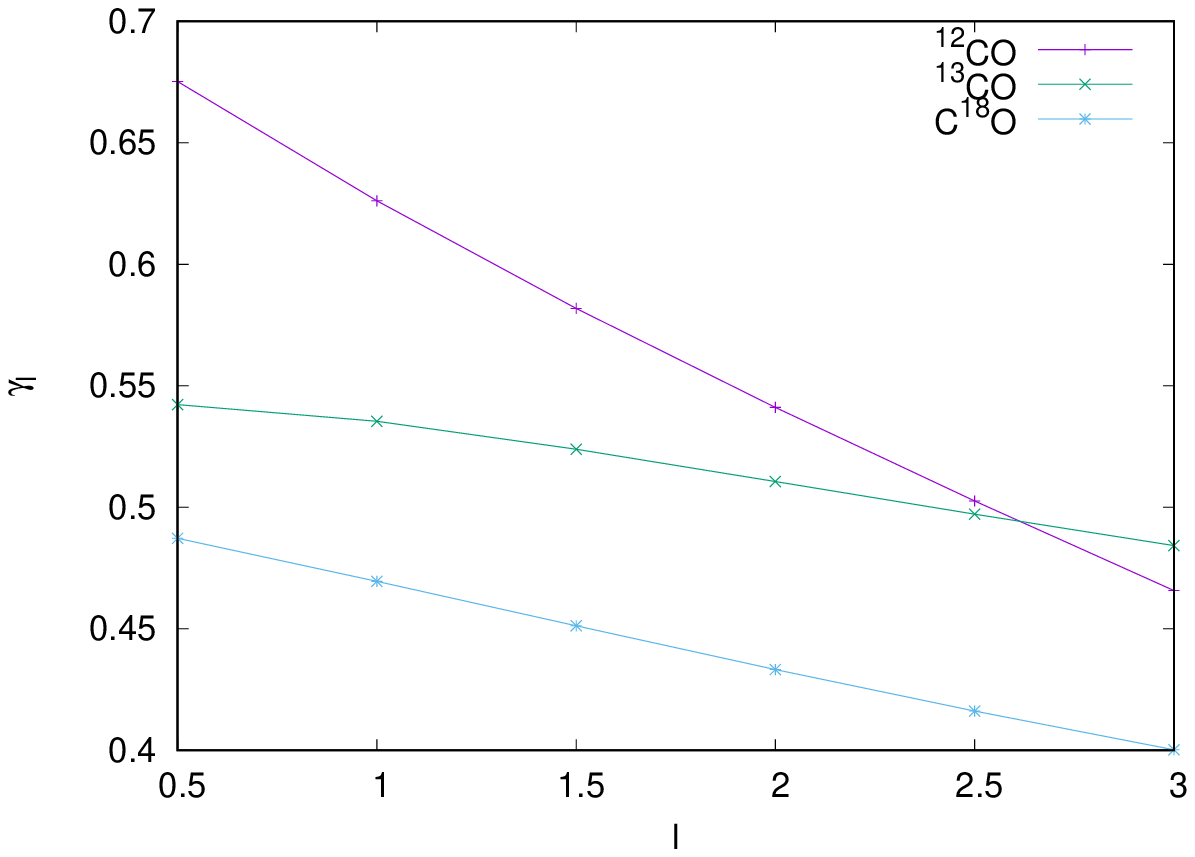} \\
\includegraphics[width=6.4cm]{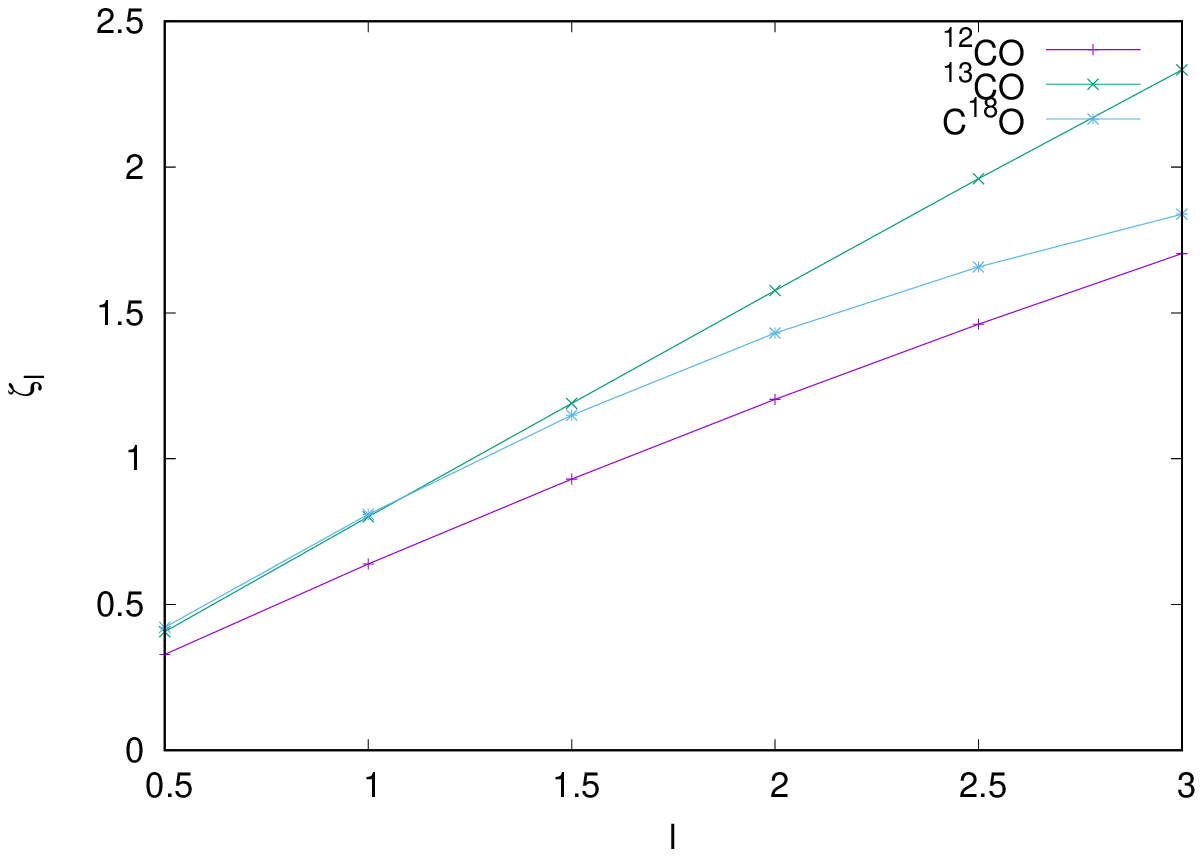} & \includegraphics[width=6.4cm]{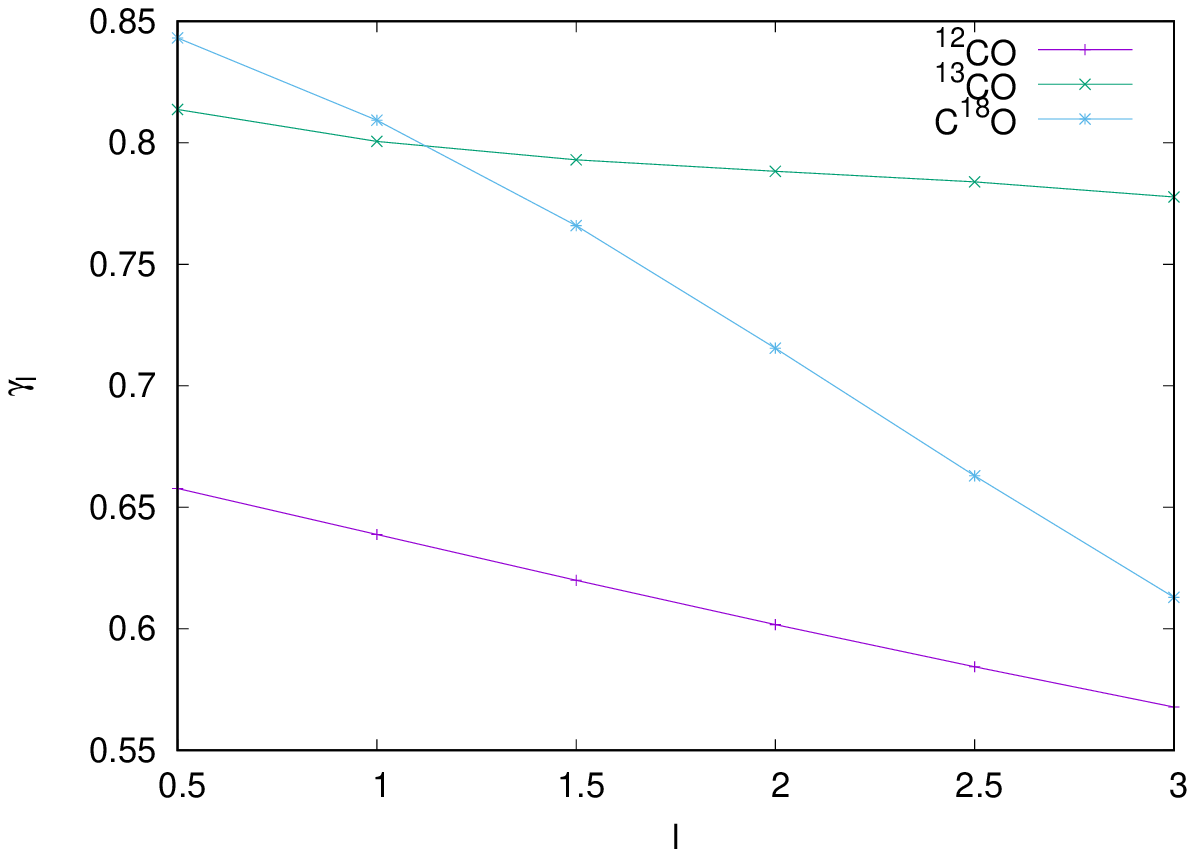} \\
\includegraphics[width=6.4cm]{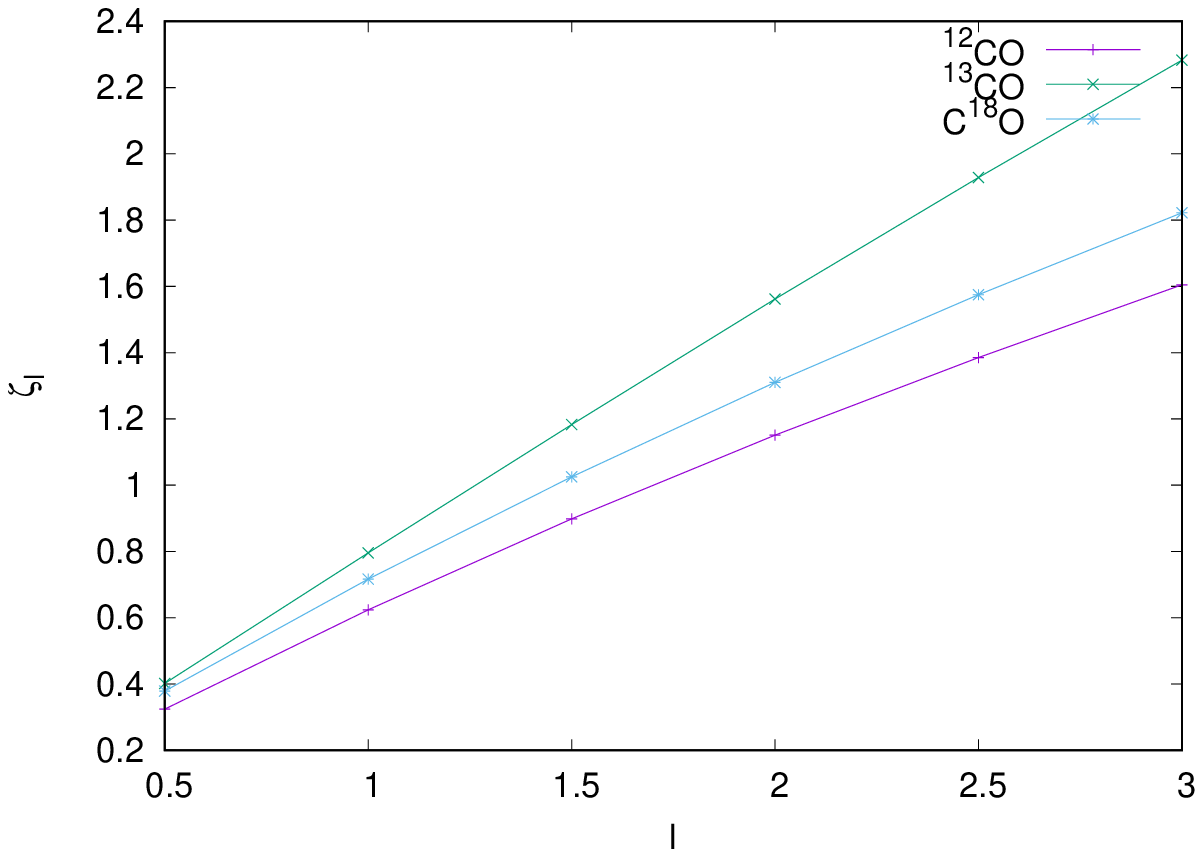}  & \includegraphics[width=6.4cm]{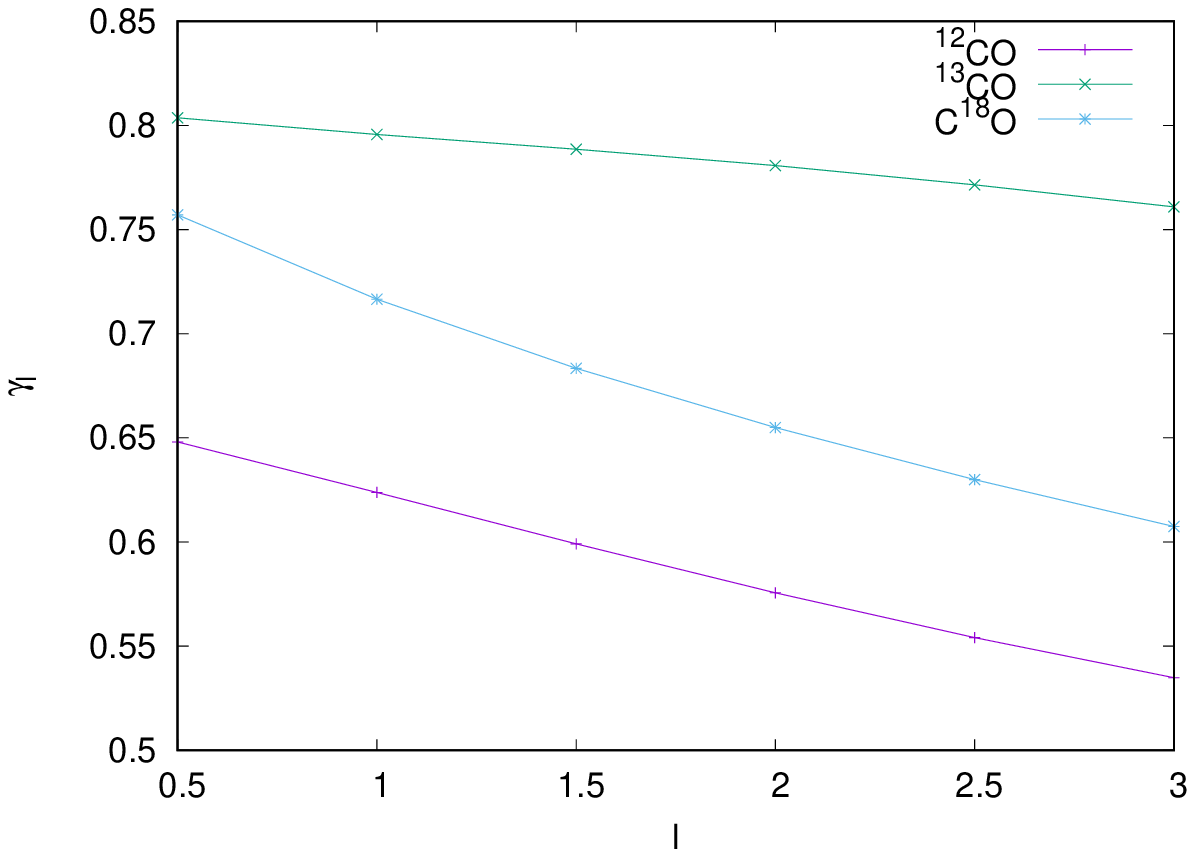} \\
\includegraphics[width=6.4cm]{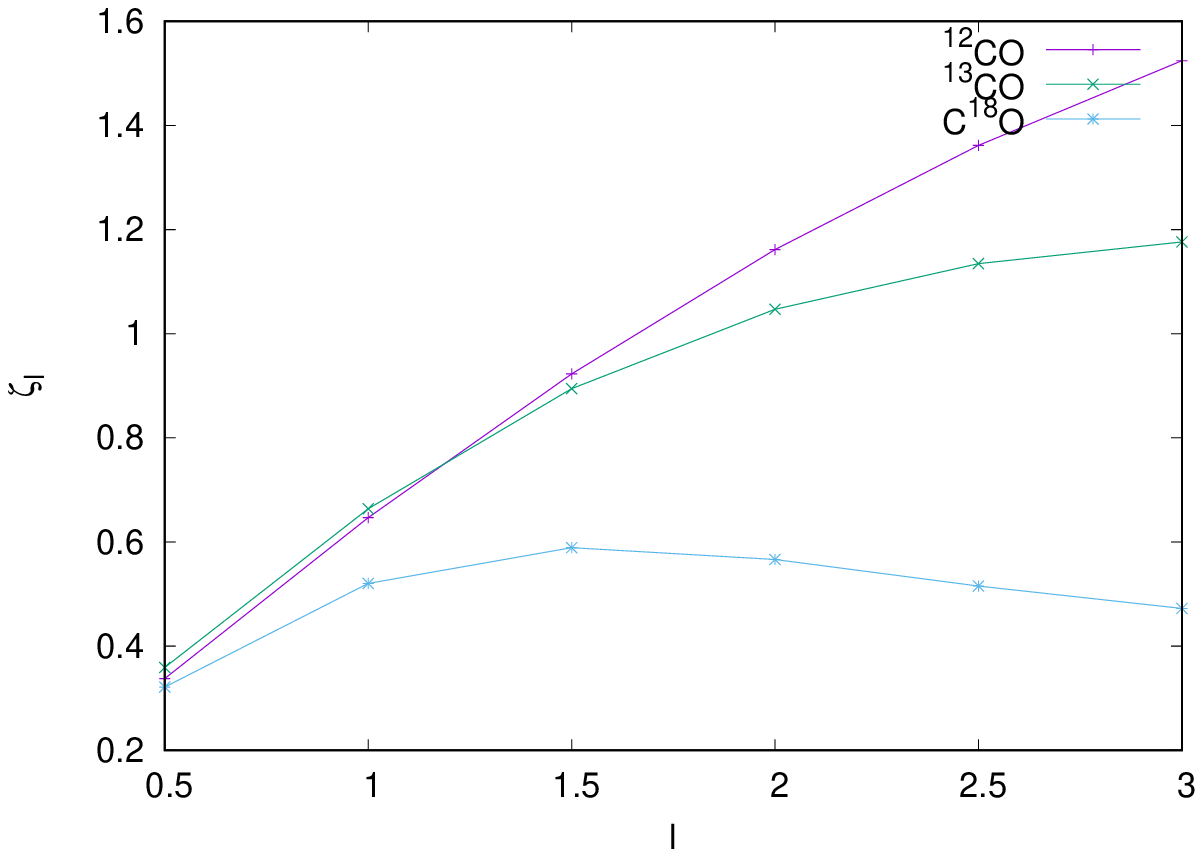} & \includegraphics[width=6.4cm]{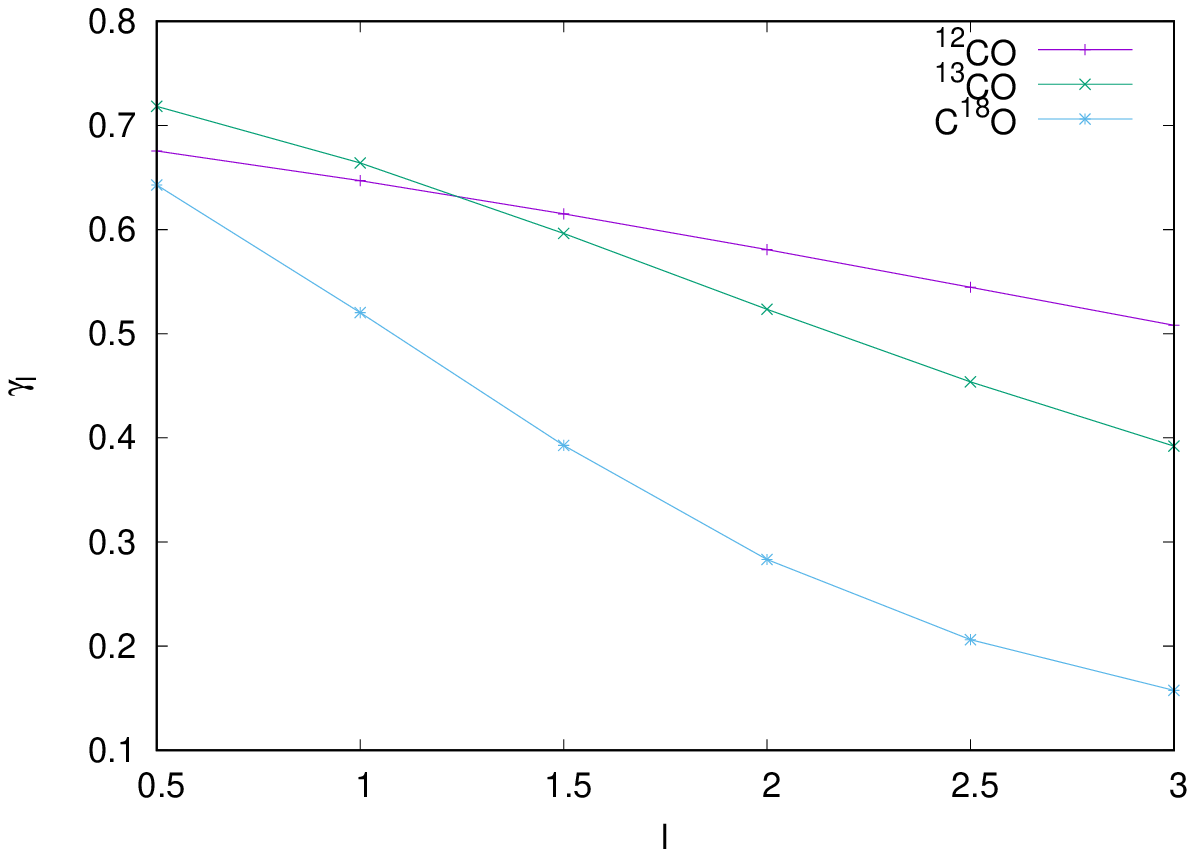} \\
\includegraphics[width=6.4cm]{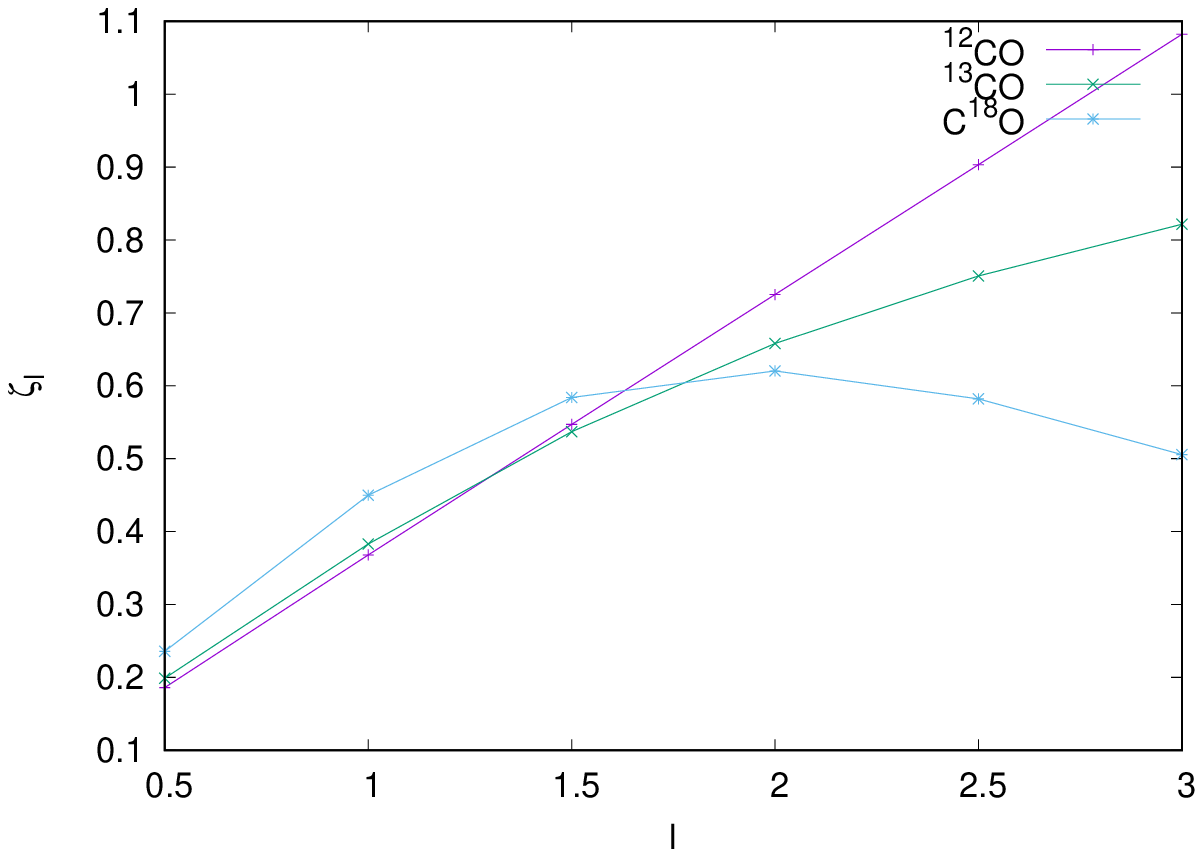}  & \includegraphics[width=6.4cm]{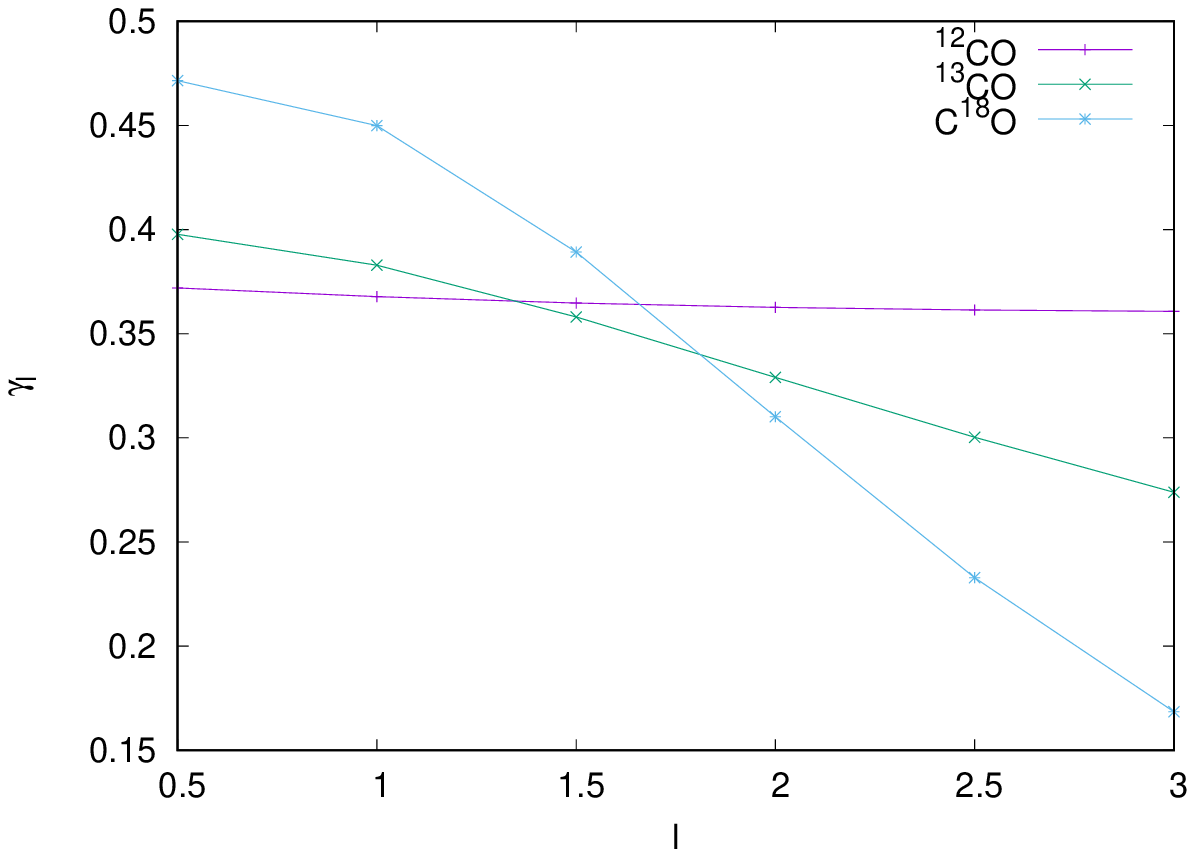} 
\end{tabular}
\caption{\label{fig:zeta-gamma} Plots of the exponents of the structure functions as a function of the order of 
the structure function, $\zeta_l$ (left column) and $\gamma_l$ (right column) for each of $^{12}$CO, $^{13}$CO and C$^{18}$O. The plots are for 
Boxes 1 -- 3, and Box 4 fitted over [0.01,0.03]\,pc and over [0.03,0.55]\,pc (top to bottom).}

\end{figure*}

\subsubsection{Implications of the structure functions for the turbulence}

We now consider the behaviour of $\zeta_l$ with $l$, which can give insight into the nature of the turbulence present. 
Figure \ref{fig:zeta-gamma} contains plots of $\zeta_l$, and of $\gamma_l \equiv \zeta_l/l$.  It is worthy of note that, for Box 4,
$\zeta_l$ is a nearly linear function of $l$ for \twco and \thco, while for \ceio it is considerably more concave. Thus, not only do our data cubes display a 
difference in the velocities detected by each line, but also in the behaviour of those velocities with distance. 
Nonetheless, all isotopes yield concave, or at most linear, behaviour for $\zeta_l$ and this is indicative of intermittency in the turbulence, 
as would be expected for molecular cloud turbulence particularly where that turbulence is supersonic.  

We can further investigate the nature of the turbulence in the Western Wall by noting that we expect that
\begin{equation}
\gamma_2 = \frac{\beta -1}{2},
\end{equation}
where $\beta$ is derived from the second order structure function. If the turbulence were incompressible then this would be an approximation of the exponent for the
kinetic energy spectrum
\begin{equation}
E(k) \propto k^{-\beta},
\end{equation}
where $k$ is the wave number. However, we must recall that our observations only give us access to the projected (line of sight) velocity, and thus this is not the true 
kinetic energy exponent.  For standard Kolmogorov turbulence $\beta = 5/3$, while for shock-dominated turbulence 
$\beta = 2$. Table \ref{table:numbers} gives the implied values of $\beta$ from our data cubes. Both \twco and \thco are
indicative of turbulence dominated by shocks, while \ceio indicates something closer to Kolmogorov-type turbulence. While it is tempting to speculate that
\ceio is tracing sub-sonic parts of the turbulent flow, this is not borne out by the behaviour of the structure functions (Fig.\ \ref{fig:structure_fns}) which
indicate this is not the case. Bearing in mind that the \ceio analysis is somewhat affected by the masking level used, as well as the concave nature of 
$\zeta_l$, we suggest that the turbulence in the Western Wall is predominantly supersonic with energy transfer between scales mediated
primarily by shocks.


\subsection{Emission Length-scales}
\label{sec:emission-length-scales}

\begin{figure*}
\centering
\begin{tabular}{cc}
\includegraphics[width=7.4cm]{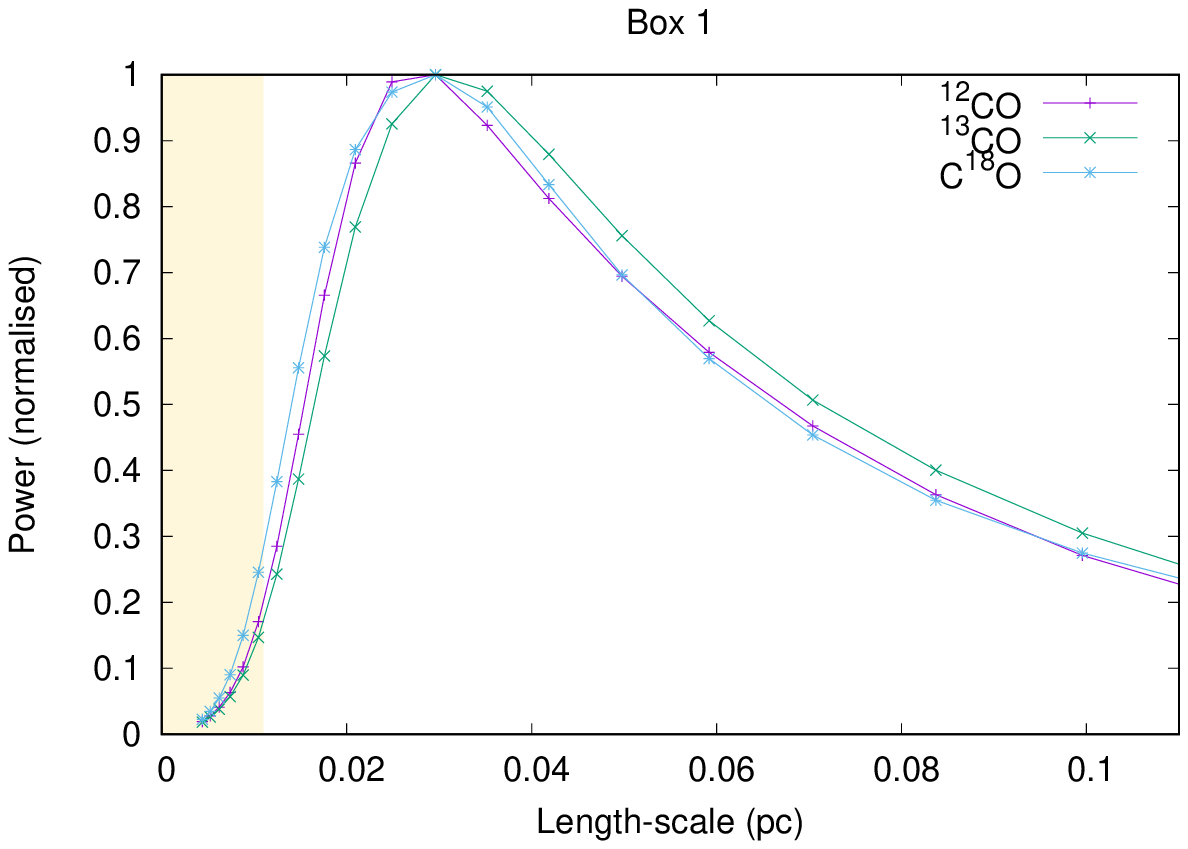} & \includegraphics[width=7.4cm]{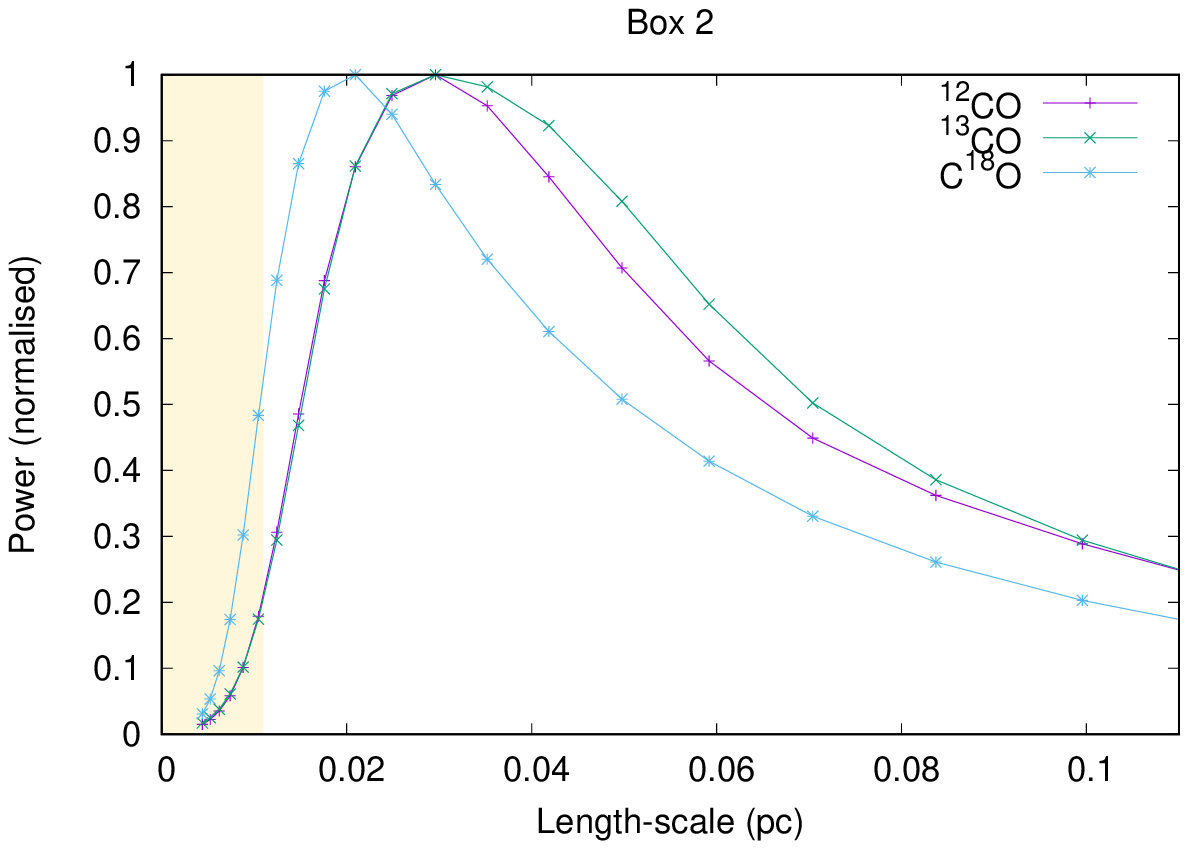} \\
\includegraphics[width=7.4cm]{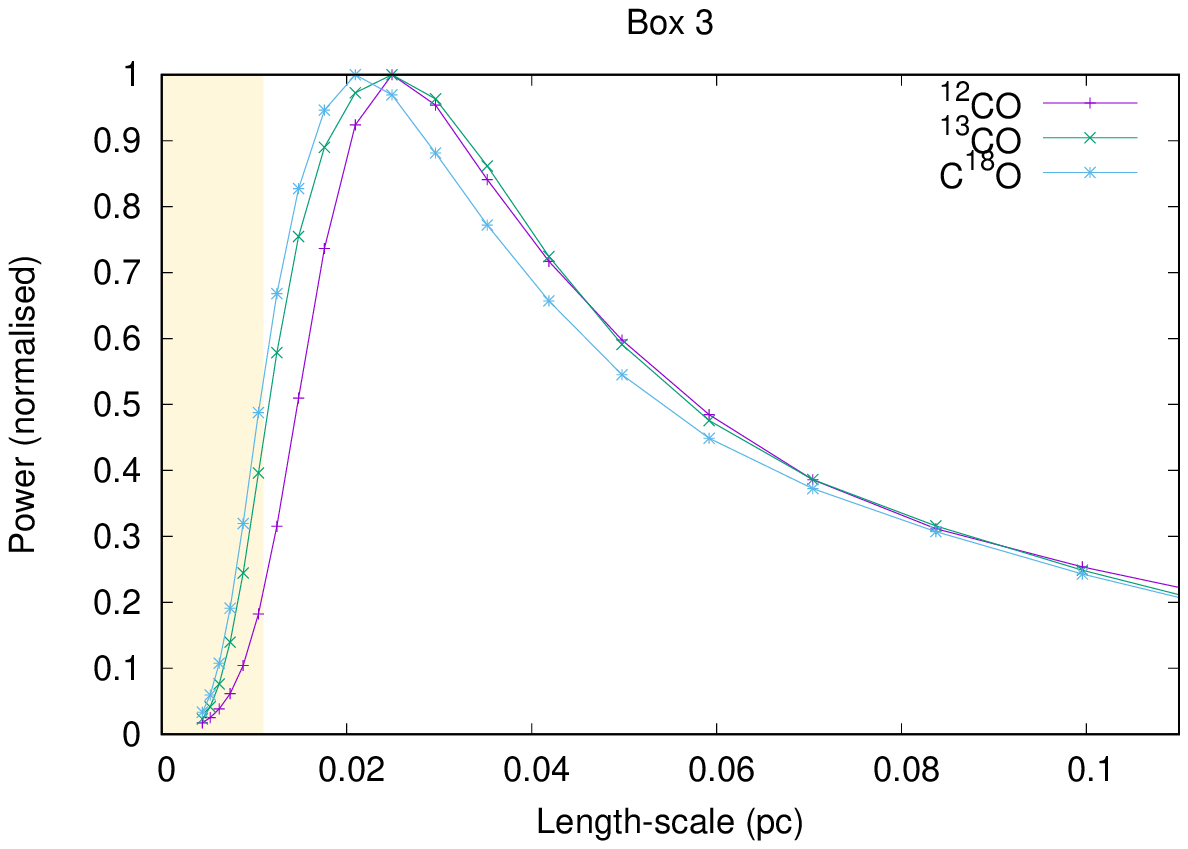} & \includegraphics[width=7.4cm]{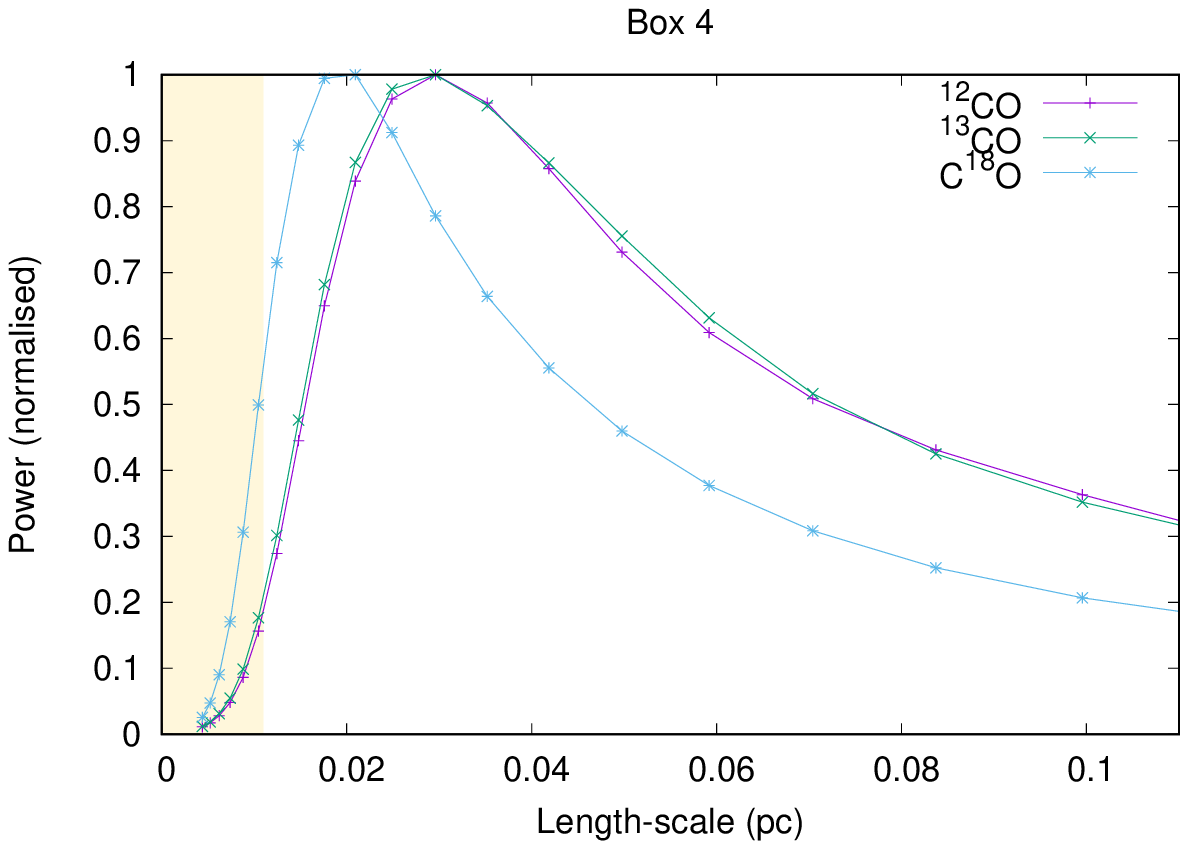}
\end{tabular}
\caption{\label{fig:spatial-spectra} Plots of the power in the variation of emission for each of the \twco, \thco and \ceio for Boxes 1 -- 4.
In each case the plots are normalised to maximum for that isotope. The shaded region in the plot indicates unresolved length-scales.}
\end{figure*}

Figure \ref{fig:spatial-spectra} contains plots of the power spectra calculated as described in Sect.\ 
\ref{sec:emission-lengthscale-analysis}. The peak power in the variation of the emission occurs at a similar
wavelength for each isotope: 0.02 -- 0.03 \,pc. This length-scale is well resolved in the observations and thus
we treat it as a physical result. It is interesting to note that the \ceio emission appears to exhibit structure on slightly shorter length-scales than \twco or
\thco in all but Box 1. This is supported by the shorter auto-correlation lengths in the PCA eigenimages for \ceio (Fig. \ref{fig:eigenimages}). In any case, the
length at which peak power occurs is close to the median of the effective core radius given by \citet{hartigan2022} (although note that in our analysis we consider 
each velocity channel independently) the sonic scale identified in Sect.\ \ref{sec:structure-fn-results}, and the location of the breaks in the centroid velocity 
structure functions. Therefore we find that a wide variety of analyses all indicate that a length-scale between 0.02\,pc and 0.03\,pc is important in the dynamics 
of the Western Wall. Finally we note that \citet{hartigan2020} discovered ``ridges'' and waves of emission in H$_2$ length-scales of approximately 0.01\,pc and 
0.025\,pc, respectively. Based on our analyses, we suggest that these ``waves'' are indicative of a population of underlying structures in the molecular cloud which 
are gradually being revealed as the photo-ionisation progresses.

\section{Conclusions}
\label{sec:conclusion}

We have performed a variety of analyses on ALMA observations of the Western Wall for \twco, \thco and \ceio. Using PCA analysis we obtain a 
large value for the ratios of the auto-correlation lengths of the first two eigenimages, suggesting that turbulence here is driven
at length-scales larger than our field of view. Combined with estimates of the sound crossing time for our analysed fields, we suggest that our
data is consistent with the turbulence being driven by processes other than irradiation by nearby massive stars. This result is in contrast with the analysis of the
pillars in the Carina Nebular presented in \cite{menonetal_21}. We do not believe there is tension between our results and those for the pillars: they are quite different 
systems with the molecular material making up the pillars clearly having been strongly sculpted for a dynamically significant amount of time. This opens up possibilities for different physical effects of the ionising
radiation. For example, \cite{lim_03} found that ionising radiation pushes a partially shadowed clump of the ISM further into the shadowed region raising
the possibility that the pillars may actually have been created partially through compressive motions initiated by the radiation field.  One expects, therefore, that the 
motions within the Pillars are likely to be strongly influenced by the process of irradiation. Such arguments, however, cannot be applied to the Western Wall where
processes such as partial shadowing of clumps is unlikely to play a significant role in the interior of the cloud.

An analysis of the centroid velocity structure functions provides more detailed insight into the dynamics of the Western Wall. We 
find some differences in the exponents of these structure functions measured using emission from the different CO isotopes, and of 
particular note is the \ceio structure function which differs markedly from both \twco and \thco. The \ceio structure functions are strongly 
influenced by our masking procedure, whereby we ignore pixels in which the data is less than 5 times the RMS noise of the data, while the 
functions for \twco and \thco are not. This is likely due to the more localised natures of the \ceio emission towards the surface of
the Western Wall. We thus proceed cautiously with regard to any conclusions based on the \ceio structure functions. The exponents 
measured for \twco and \thco indicate that the region is dominated by supersonic turbulence, while the data for \ceio suggests the turbulence is 
sub-sonic, with a kinetic energy spectrum expected of Kolmogorov turbulence. Based on the masking issue discussed above, and the concave nature of $\zeta_l$ we 
favour concluding that the region is indeed dominated by supersonic turbulence in which much of the kinetic energy dissipation occurs via shocks.

The first order centroid velocity structure functions can be used to measure the sonic scale of the turbulence, and this is 
inferred to have an upper limit of between 0.01 and 0.02\,pc in the Western Wall. This is comparable to the sonic scale found in, for example,
the simulations of \citet[see Fig.\ 14 of][]{downes12} and thus suggests that the turbulence, while supersonic, may not be strongly so. 
Indeed, it is worth noting that the maximum value of the first order structure functions corresponds to a Mach number of around 2 or 3.

\citet{hartigan2022} note a 2\kms South to North gradient in the mean centroid velocity measured in \twco across the entire observed 
field. This may lead to an over-estimate of the turbulent velocities and thus an under-estimate of the sonic scale. However, the gradient in our data is very modest, 
keeping in mind that we analyse subsets of the data and the centroid velocity structure functions are calculated for distances up to roughly 6\% of the size of our observational field in Box 4. Performing the analyses with and without 
gradient subtraction demonstrates that this does not significantly impact our 
results, and this conclusion holds for any reasonable East to West gradient 
subtraction also.

A further analysis of the data, using power spectra of the spatial variations in the emission, indicates that there is a preferred
length-scale of 0.02 -- 0.03\,pc for these variations. We note that this corresponds approximately to the median clump size
quoted in \citet{hartigan2022}, and is also of a similar scale to the sonic scale of the turbulence and the location of breaks in the
structure functions. Furthermore, this length-scale is the same as that of the ``waves'' reported by \cite{hartigan2020} in their observations of
H$_2$ emission. This is clear evidence that the internal structure of an irradiated molecular cloud can be studied by examining the
geometry of the irradiate surface.

Interestingly, \citet{federrath16a} report simulations which exhibited sonic scales in the range 0.04 -- 0.16\,pc, of
order the typical filament widths in molecular cloud turbulence \citep{arzoumanianetal_11, ntormousi16}, and considerably {\em larger} 
than those measured here or in \citet{downes12}. \citet{federrath16a} postulates a link between the sonic scale and the filament widths 
observed in molecular clouds. Although the actual length-scales are different, this link itself is supported by our analysis for the 
Western Wall since the sonic scale is of order the filament width and, indeed, it is of order the dominant length-scale generally 
(whether filaments or clumps) in the observations.

This apparently significant length-scale is well below the length-scales which we would expect to be strongly affected by 
ambipolar diffusion, and so it seems unlikely that the turbulence cascade itself is directly producing structures at this length-scale. We 
therefore suggest that these 
structures occur as a result of the effects of gravity. This does not imply that structures on this scale are gravitationally bound, but rather that the action of gravity, 
in combination with turbulence, is likely to be responsible for their formation. It is well known that supersonic isothermal turbulence should result in a log-normal
probability density function for the mass density, and that if self-gravity is included then this log-normal distribution gains an
exponential tail at high densities \citep{federrath_13}. Our interpretation of these observations suggest that the dominant length-scale we find is related
to structures in this exponential tail.

\section*{Acknowledgements}
A. I. acknowledges support from the ALMA Study Project \#358232. A. I. and M. H. acknowledge support from
the National Science Foundation under grant No. AST-1715719. This paper makes use of the following ALMA
data: ADS/JAO.ALMA\#2015.1.00656.S. ALMA is a partnership of ESO (representing its member states),
NSF (USA) and NINS (Japan), together with NRC (Canada), MOST and ASIAA (Taiwan), and KASI (Re-
public of Korea), in cooperation with the Republic of Chile. The Joint ALMA Observatory is operated by
ESO, AUI/NRAO and NAOJ. The National Radio Astronomy Observatory is a facility of the National 
Science Foundation operated under cooperative agreement by Associated Universities, Inc.

\section*{Data Availability}
The data underlying this article will be shared on reasonable request to the corresponding author.



\bibliographystyle{mnras}
\bibliography{final_paper} 



\bsp	
\label{lastpage}
\end{document}